\documentclass[pdflatex,sn-mathphys-num]{sn-jnl}

\usepackage[utf8]{inputenc}
\usepackage{graphicx}%
\usepackage{multirow}%
\usepackage{amsmath,amssymb,amsfonts}%
\usepackage{amsthm}%
\usepackage{mathrsfs}%
\usepackage[title]{appendix}%
\usepackage{xcolor}%
\usepackage{textcomp}%
\usepackage{manyfoot}%
\usepackage{booktabs}%
\usepackage{algorithm}%
\usepackage{algorithmicx}%
\usepackage{algpseudocode}%
\usepackage{listings}%
\usepackage{fullpage}

\usepackage{array}
\usepackage{isotope}
\usepackage{bm}
\usepackage{orcidlink}

\usepackage[final]{changes}

\newcommand{\cs}{\isotope[137]{Cs}}
\newcommand{\tb}{\isotope[161]{Tb}}
\newcommand{\Lu}{\isotope[177]{Lu}}
\newcommand{\lu}{\isotope[177]{Lu}}

\theoremstyle{thmstyleone}%
\theoremstyle{thmstyletwo}%

\theoremstyle{thmstylethree}%

\begin{document}

\title[Article Title]{Multicenter Comparison of Radionuclide Calibrators and SPECT/CT Protocols for Quantitative \Lu \, Imaging in Clinical Practice}

\author*[1]{\fnm{Wies}\sur{Claeys}\orcidlink{0009-0002-2616-7119}}\email{wies.claeys@kuleuven.be}
\author[1]{\fnm{Kristof} \sur{Baete} \orcidlink{0000-0003-0113-1590}}
\author[2]{\fnm{Laurence} \sur{Beels}\orcidlink{0000-0002-6999-9575}}
\author[3]{\fnm{Claire} \sur{Bernard} \orcidlink{0009-0001-6574-5164}}
\author[4,5]{\fnm{Rachele} \sur{Danieli} \orcidlink{0000-0001-8013-4938}}
\author[6]{\fnm{Yves} \sur{D’Asseler}\orcidlink{0000-0002-8807-2257}}
\author[7]{\fnm{An} \sur{De Crop}}
\author[8]{\fnm{Michel} \sur{Hesse} \orcidlink{0000-0002-9434-751X}}
\author[1,9]{\fnm{Victor} \sur{Nuttens} \orcidlink{0000-0002-6797-8789}}
\author[4,5]{\fnm{Bruno} \sur{Vanderlinden}}
\author[1]{\fnm{Michel} \sur{Koole} \orcidlink{0000-0001-5862-640X}}

\affil*[1]{\orgdiv{Nuclear Medicine and Molecular Imaging}, \orgname{KU Leuven},  \city{Leuven}, \country{Belgium}}
\affil[2]{\orgdiv{Nuclear Medicine}, \orgname{AZ Groeninge}, \city{Kortrijk}, \country{Belgium}}
\affil[3]{\orgdiv{Nuclear Medicine}, \orgname{Centre Hospitalier Universitaire de Liège}, \city{Liège}, \country{Belgium}}
\affil[4]{\orgdiv{Medical Physics}, \orgname{Institut Jules Bordet, Université Libre de Bruxelles (ULB), Hôpital Universitaire de Bruxelles (H.U.B)}, \city{Brussels}, \country{Belgium}}
\affil[5]{\orgdiv{Radiophysics and MRI Physics Laboratory}, \orgname{Université Libre de Bruxelles (ULB)}, \city{Brussels}, \country{Belgium}}

\affil[6]{\orgdiv{Nuclear Medicine}, \orgname{Ghent University Hospital}, \city{Ghent}, \country{Belgium}}
\affil[7]{\orgdiv{Nuclear Medicine}, \orgname{AZ Delta}, \city{Roeselare}, \country{Belgium}}
\affil[8]{\orgdiv{Nuclear Medicine}, \orgname{Cliniques Universitaires Saint-Luc}, \city{Brussels}, \country{Belgium}}
\affil[9]{\orgdiv{Nuclear Medicine}, \orgname{AZORG}, \city{Aalst}, \country{Belgium}}

\abstract{
\begingroup \unboldmath
\textbf{Purpose:} Following the clinical success of \lu-based therapies for neuroendocrine tumors and prostate cancer, accurate quantification of \lu\, using radionuclide calibrators (RNCs) and \lu-SPECT/CT is gaining importance as prerequisite for accurate treatment delivery and dosimetry. However, the lack of standardization can introduce inter-system variability, compromising multi-center clinical trials. This study aimed to assess the accuracy and variability of \lu\, measurements using RNCs and SPECT/CT across different systems and hospitals.

\textbf{Methods:} A uniform cylindrical phantom and a NEMA phantom with hot spheres were prepared using traceable activities and imaged at 8 different hospitals using 13 SPECT/CT systems (9 conventional and 4 3D CZT). Acquisitions and reconstructions were performed using both site-specific and standardized protocols. The cylindrical phantom images were used to \added{evaluate the system calibration and} establish image calibration factors (ICFs), the NEMA images to evaluate \replaced{effective resolution}{image quality} by calculating recovery coefficients (RCs).\deleted{Both were used to evaluate quantitative SPECT accuracy.} In parallel, two vials were measured to test RNC accuracy.

\textbf{Results:} RNC measurements differed up to 11\% between centers, while SPECT quantification \added{of the cylindrical phantom} differed up to 20\% \deleted{and 58\% for the cylindrical and NEMA phantoms respectively}. While ICFs were consistent for systems of the same type, image quality varied strongly when using clinical protocols\added{ (36\% difference in RCs in the largest sphere)}. Standardized reconstruction reduced variability \added{in RCs} for each system type \added{(maximum 12\% difference)}, regardless of acquisition protocol, but differences between system types persisted when \replaced{standardizing}{harmonizing} acquisition and reconstruction \added{(33\% difference)}.

\textbf{Conclusion:} Current \lu \,measurement practices yield significant variability in quantification and image quality. Harmonization efforts should prioritize standardized calibration and reconstruction protocols to improve multicenter reproducibility of quantitative \lu-SPECT/CT. \endgroup}


\maketitle

\section{Introduction}

Quantitative single-photon emission computed tomography (SPECT) combined with \added{computed tomography (}CT\added{)} has emerged as an essential imaging modality in targeted radionuclide therapy (TRT), particularly for treatments using lutetium-177 (\lu). The rapidly growing clinical implementation of \lu-PSMA therapy in metastatic castration-resistant prostate cancer \cite{LuPSMA1,LuPSMA2,LuPSMA3} and \lu-DOTATATE therapy in neuroendocrine tumors \cite{LuDOTA1,LuDOTA2} has accelerated the demand for reliable post-therapy SPECT imaging. In particular, \lu-SPECT is increasingly used to quantify activity in target regions and organs at risk, enabling patient-specific dosimetry, and individualized treatment planning \cite{EANM1,PractKidneyDosim}. Beyond treatment planning, reliable dosimetry provides a framework for improving our understanding of dose–effect relationships, validating imaging biomarkers, and optimizing therapy in a multi-center setting, thereby strengthening the evidence base and regulatory efforts toward more personalized nuclear medicine.

While quantitative SPECT remains more technically challenging compared to \added{positron emission tomography (}PET\added{)} \added{due to the poorer spatial resolution and sensitiviy and the wide range of relevant photon energies}, current SPECT/CT technology, when properly calibrated and standardized, can achieve quantitative accuracy that is sufficient for clinical dosimetry in \lu-based therapies \cite{qSPECT1,qSPECT2,MIRD23,MIRD26}. However, several recent phantom and multi-center studies have found large variability in quantification and \replaced{recovery coefficients}{image quality} between different hospitals \cite{Wevrett,SteffieP}, while other have achieved harmonized performance by \replaced{standardizing}{harmonizing} imaging protocols \cite{MRT,SteffieP2}. \added{Similarly, it has been shown that harmonizing dosimetry workflows can reduce variability in the absorbed dose estimates \cite{Jessen}.} This underscores that reconstruction, calibration, and post-processing protocols strongly influence quantitative outcomes. Since inter-system variability can compromise multi-center trials and hinder the establishment of reproducible dosimetry frameworks, it is critical to quantify the degree of variability between systems and protocols, to inform consensus recommendations and standard operating procedures that ensure comparability between hospitals.  Initiatives such as the EANM Research Ltd. (EARL) standards have demonstrated the importance of protocol \replaced{harmonization}{standardization} and centralized accreditation to enable cross-center comparability in PET/CT \cite{EARLPET1,EARLPET2}, yet equivalent \deleted{harmonization} frameworks for \lu-SPECT/CT are still under development \cite{EARL_abstract}.

Similarly, previous studies have found large variability in activity measurements using radionuclide calibrators (RNCs) at different hospitals, due to a reliance on manufacturer settings and absence of standardized calibration procedures \cite{Clarita, Clarita2,Fenwick}. Reliable RNC measurements are not only crucial to determine the activity administered to the patient, but also for accurate dosimetry, since most hospitals use an in-house RNC to calibrate their SPECT/CT systems and any calibration uncertainty contributes to the uncertainty in the final \added{absorbed} dose estimate \cite{EANM_uncertainties}. Therefore end-to-end metrology, including traceable RNC measurements, are a prerequisite for accurate quantitative SPECT imaging and dosimetry \cite{prismap}\added{\cite{Robinson}}.

Meanwhile, the introduction of solid-state 3D cadmium zinc telluride (CZT)–based SPECT/CT systems has created new opportunities for quantitative imaging. These systems offer higher sensitivity, lower susceptibility to dead-time effects, improved energy resolution and novel detector geometries compared to conventional NaI(Tl) systems, enabling shorter acquisition times and potentially more precise activity recovery \cite{CZT1,CZT2}. Early reports demonstrate that CZT SPECT/CT can achieve robust quantitative accuracy for \lu\, and may outperform conventional systems in lesion detectability and dosimetry reproducibility, while simultaneously reducing scan times in post-therapy imaging \cite{CZT3,CZT4}. As such, 3D CZT systems are expected to play an increasingly important role in clinical trials and routine therapy monitoring. However, a comprehensive comparison between the quantitative capabilities of 3D CZT and conventional SPECT/CT systems is still lacking.

\replaced{The aim of this research was to assess the current variability in RNC measurements of \lu \, and quantitative \lu-SPECT between different hospitals, and to study the impact of standardizing reconstruction and acquisition protocols on the inter-center variability. Both conventional and 3D CZT systems at academic and non-academic centers were included to obtain a representative sample of clinically used systems.}{The present study systematically compares the quantitative \lu\, performance of RNCs and SPECT/CT scanners of different vendors across multiple centers, including conventional and 3D CZT systems at academic and non-academic institutions. Vials with traceable activities are used to evaluate RNCs, and standardized phantom experiments are used to evaluate SPECT quantification, sensitivity and image quality. Measurements are performed using existing clinical protocols as well as standardized protocols to assess both the current clinical situation and the potential of standardization.} The findings are intended to support harmonization efforts and contribute towards establishing \replaced{accreditation}{vendor-independent} standards for quantitative \lu-SPECT/CT imaging in both clinical practice and multi-center research.

\section{Materials and Methods}
Two vials and two phantoms were prepared \replaced{with traceable \lu \, activities}{and measured} at UZ Leuven, similarly to an earlier calibration effort for \tb \, \cite{NMES}, before being shipped for measurement at seven other Belgian hospitals, including AZ Delta, AZ Groeninge, AZORG Aalst, CHU de Liege, Cliniques Universitaires Saint-Luc, Institut Jules Bordet and UZ Gent. Participating sites were selected to provide a representative sample of the most common SPECT/CT scanners, including 3D CZT systems. Furthermore, eligible centres were required to routinely perform \lu-therapies with post-therapy SPECT imaging. \added{The vials were used to evaluate the RNCs each site, and phantom experiments were used to evaluate SPECT quantification, sensitivity and image quality. Measurements were performed using existing clinical protocols as well as standardized protocols to assess both the current clinical situation and the impact of protocol standardization.} Because of the half-life of \lu, the measurement itinerary was constrained to a two-week window, allowing two days per site to perform the requested measurements.

\subsection{Vial and Phantom Preparation}
Two vials with different geometries were used to test the accuracy of the RNCs at the different clinical sites. A SCHOTT 10R type 1+ vial (SCHOTT Pharma) containing 4 g of solution \deleted{and a nominal activity of 300 MBq} was used as the reference geometry, as it is \replaced{commonly}{typically} used in \added{traceable} RNC calibrations. In addition, an SV-25A vial (Huayi Isotopes Co.) containing 10 g of solution \deleted{and a nominal activity of 150 MBq} was used to represent the typical clinical vial geometry.

\replaced{These}{Both} vials were filled \replaced{with 300 MBq and 150 MBq }{from a stock solution} of carrier-free \lu\, \replaced{chloride}{Cl$_2$} (ITM Isotope Technologies Munich) \added{respectively}, with EDTA added to avoid adsorption to the vial walls and to ensure a homogeneous solution. The reference activity for both vials was determined using two reference RNCs (a VIK-202 (COMECER S.p.A.) and a CRC-55tR (Capintec, Mirion Technologies, Inc.) that were previously calibrated against a Fidelis (Southern Scientific Ltd.) secondary standard RNC owned and maintained by the Laboratory for Nuclear Calibrations (LNK) at the Belgian Study Center for Nuclear Energy (SCK CEN). \added{During this calibration, three reference vials were measured using the Fidelis and each reference RNCs to obtain a calibration factor, from which a new dial settings was derived, as detailed in appendix \ref{app:calibration_details}.} \replaced{Afterwards, }{Since this calibration was performed for the reference geometry only, }an additional correction factor was determined to account for the different geometry of the clinical vial.

The uncertainty on the reference activity in both vials was estimated following the National Physical Laboratory (NPL) Good Practice Guide No. 93 \cite{gpg}. The uncertainty budget included statistical uncertainty, vial positioning, constancy, linearity and geometry effects (uncertainty in the geometry correction factor and deviations from the nominal volume).

For SPECT imaging, a NEMA IEC body phantom with 6 hot spheres (diameters: 13 - 60 mm, nominal activity concentration: 4 MBq/mL) and a cold background compartment (no lung insert) was used alongside a uniform cylindrical phantom (internal length: 211 mm, internal diameter: 195 mm, \added{volume: 6.3 l, }nominal activity: 800 MBq). Again, a solution of \lu\, \replaced{chloride}{Cl$_2$} with EDTA was used to ensure that the activity distribution remained homogeneous over time.

After filling the phantoms, the \replaced{exact}{effective} activity was determined using gamma counter measurements. Three 1 mL samples each were taken from the cylindrical phantom and the stock solution used to fill the spheres of the NEMA phantom. Using a Wizard 1480 gamma counter (PerkinElmer) that had been cross-calibrated against the two reference RNCs, the activity concentration in these samples was measured. The total activity in each phantom was then calculated by multiplying the measured activity concentration by the phantom volume, which had been determined beforehand by filling with a known volume of water. To verify that the activity concentration remained constant throughout the measurement campaign, additional gamma counter samples were taken from the phantoms after the completion of all SPECT measurements. The uncertainty on the phantom activities was estimated by considering statistical counting uncertainty, constancy and phantom volume, in addition to the components included in the uncertainty budget for the vial activities.

More details on the calibration procedures can be found in Appendix \ref{app:calibration_details}.

\subsection{Radionuclide Calibrators}

To assess the accuracy of routine RNC measurements across the different sites, each site measured the two vials using its routine clinical RNC procedure (i.e. using any custom dial settings, correction factors and positioning aids). Additional measurements using factory protocol (i.e. default dial settings, no correction factors and using only the standard dipper and liner) were performed to assess variations between RNCs when using identical measurement protocols.

To mitigate variations in activity readings due to noise, electrometer fluctuations and differences in vial positioning, at least 9 measurements per vial and per protocol were acquired over several hours, repositioning the vial between each reading. Measurements were background corrected, decay corrected to a common reference time and averaged.

Additionally, a questionnaire was distributed to collect information on routine quality controls and custom measurement and calibration procedures, thereby enabling a detailed comparison of the RNCs across sites.

\subsection{SPECT/CT}

All sites performed SPECT/CT acquisitions of the two phantoms and reconstructed the data using their routine clinical procedures, in addition to acquisitions and reconstructions using protocols with varying levels of standardization. In total, three scenarios were defined: non-standardized (acquisition and reconstruction following the normal clinical protocols), semi-standardized (acquisition following the clinical protocol, reconstruction using standardized settings) and fully standardized (acquisition and reconstruction following a standardized protocol).

The standardized acquisition and reconstruction parameters were based on the preliminary EARL SPECT protocol \cite{EARL_param} and are listed in tables \ref{tab:acq_param} and \ref{tab:recon_param}. These settings are defined for conventional SPECT systems, and cannot be directly translated to CZT systems because of inherent differences in system design, such as the 3D acquisition mode, integrated collimation and different energy ranges. Consequently, no standardized acquisition settings were defined for CZT systems. The measurement time was adapted to reach a fixed number of counts per acquisition, in order to obtain similar count statistics across all images, regardless of system sensitivity and the physical decay of the activity during the measurement campaign.

Note that these standardized parameters cover the settings that are most commonly understood to have a large impact on the resulting image and that can be readily adapted in most reconstruction software. However, some vendors allow the user to change additional settings, potentially introducing extra variability between systems. In particular, it was found during the study that some Siemens reconstructions used 20 mm Gaussian smoothing on the scatter windows before reconstruction, while others did not apply any pre-filtering. Further investigation (see Appendix \ref{app:scatter_window_smoothing}) showed that filtering the scatter windows has a beneficial effect on the reconstructed image, in line with previous research \cite{scatter_methods}\cite{scatter_smoothing}. Therefore, the standardized reconstructions of the Siemens data were repeated using a \replaced{standardized}{harmonized} 20 mm Gaussian filter on the scatter windows. For the other vendors, the default settings (no pre-filtering of the scatter window) were retained.

To evaluate \added{how errors in system calibration impact} the accuracy of the SPECT quantification, sites were asked to estimate the total activity inside each phantom from the images acquired with their routine clinical protocols using their existing calibration factors. Sites were informed of the total sphere volume (161 mL) and the volume of the cylindrical phantom (6.3 L), but were free to use the calculation methods of their own choice. The reported activities were compared to the reference activity obtained from the gamma counter measurements.
Finally, a questionnaire was sent to all sites to collect information on the clinical protocols and on calibration and quality control procedures.

\begin{table}
    \centering
    \caption{Parameters of the standardized acquisition protocol}
    \label{tab:acq_param}
    \begin{tabular}{c|c}
    Collimator & medium energy\\
    Matrix size    &  256 x 256\\
    Views     & 2 x 60 \\
    Photopeak window & 208 keV $\pm$ 10 \% \\
    Scatter windows & 10\% adjacent upper \& lower scatter \\
    Stop condition & 7 MCts (Cylinder) or 3 MCts (NEMA)
    \end{tabular}
\end{table}

\begin{table}
    \centering
    \caption{Parameters of the standardized reconstruction protocol}
    \label{tab:recon_param}
    \begin{tabular}{c|c}
    Type & Vendor recon \\
    Algorithm     &  OSEM\\
    Corrections     & AC, window-based SC, RR \\
    Updates & 40i2s or 20i4s \\
    Post-filtering & None
    \end{tabular} 
\end{table}

All SPECT images were collected and analyzed centrally. The images of the uniform cylindrical phantom were used to calculate the image calibration factor (ICF), which represents the number of counts in the reconstructed image per unit time and per unit activity. The ICF is required to quantify SPECT images and to calculate recovery coefficients (see below), and can also provide an indication of the sensitivity of conventional systems.

The ICF \added{(cps/MBq)} was calculated similarly to \cite{SteffieP}, using the counts concentration $[C]$ in a volume of interest (VOI) inside the cylinder:
\begin{equation}
    ICF = \frac{[C]}{\Delta t\, [A_{C}]} \, ,
\end{equation}
Here, $\Delta t$ is the acquisition duration and $[A_C]$ is the activity concentration in the cylinder, measured using the gamma counter. This method was implemented in Python using a cylindrical VOI of 12 cm diameter and 15 cm length, as this was found to be the largest contour that excluded the Gibbs artifacts in all of the images. 

A second approach using an expanded VOI containing all counts originating from the cylinder as in \cite{MRT} was implemented as well, and the two methods are compared in appendix \ref{app:small_versus_large_volume}. ICFs calculated with both methods differed up to $7\%$ and the second approach was found to be more susceptible to scatter, incorrect attenuation correction and septal penetration, so only the first method was used for the final analysis.

The images of the NEMA phantom were analyzed to determine the recovery coefficients (RCs) of each sphere:
\begin{equation}
   \mathrm{RC}_i = \frac{[A_i]}{[A_{N}]} \, .
\end{equation}
Here, $[A_i]$ is the activity concentration in sphere $i$ as measured on the SPECT image, and $[A_{N}]$ is the activity concentration in the NEMA phantom measured using the gamma counter.

For SPECT images with voxel values proportional to counts, the activity in the spheres was quantified using the ICF obtained from the cylinder image with the same acquisition and reconstruction parameters:
\begin{equation}
    [A_i] = \frac{[C_i]}{\mathrm{ICF}\, \Delta t} \, ,
\end{equation}
where $[C_i]$ is the mean counts concentration in sphere $i$.

\replaced{For some SPECT systems, the reconstructed images were already expressed as activity concentration (Bq/ml), so in principle $[A_i]$ could be obtained directly from the images. However, since any error in quantification translates into errors in RCs, an extra correction factor CF (MBq/MBq) was used instead to minimize bias in the recovery curves:}{For SPECT images with voxel values already expressed as activity concentration, the measured activity concentration in each sphere, $[A_{\mathrm{meas},i}]$, was corrected using a global correction factor CF to minimize potential bias in the recovery coefficients due to calibration errors:}
\begin{equation}
    [A_i] = \frac{[A_{\mathrm{meas},i}]}{\mathrm{CF}} \, .
\end{equation}
where CF was obtained from the uniform cylinder as
\begin{equation}
    \mathrm{CF} = \frac{[A_{\mathrm{meas},C}]}{[A_{C}]} \, ,
\end{equation}
where $[A_{\mathrm{meas},C}]$ is the activity concentration measured in the same VOI as in the ICF calculation described above. The counts or activity concentration in each sphere was obtained using a Python script performing a SPECT-based segmentation based on the theoretical sphere sizes.

Many of the images showed pronounced Gibbs artifacts (GAs) inside the 60 mm sphere. These were quantified similarly to \cite{Gibbs,Gibbs2}: a Python script was used to plot the radial profile of the 60 mm sphere and fit a smoothing spline to the data. This spline was used to determine the height of the Gibbs maximum $M$ and minimum $m$, and the GA strength was calculated as
\begin{equation}
    \mathrm{GA} = \frac{M - m}{M + m} \, .
\end{equation}

\section{Results}
To ensure confidentiality, sites are labeled A–H, matching the order of the measurement itinerary. Numbers are used to distinguish multiple systems at the same site.

\subsection{Vial and Phantom Preparation}

The calibration settings for the two reference RNCs are summarized in Table \ref{tab:RNCs}. The geometry correction factors for the clinical vial were $1.006\pm0.002$ and $1.004\pm0.002$ for the VIK-202 and CRC-55tR respectively. At the start of the measurements, the activity was 291 MBq in the reference vial and 158 MBq in the clinical vial. The uncertainty on both measurements was estimated at 3.0\% (k=2), which was dominated by the uncertainty on the reference activity determined by the Fidelis (2.8\%).

For the uniform cylindrical phantom, the gamma counter samples taken before and after the SPECT measurements agreed within the measurement uncertainty, showing that the activity concentration remained constant throughout the measurement campaign. The \deleted{effective} activity at the start of the measurements was 812 MBq with an uncertainty of 3.3\% (k=2). For the NEMA phantom, the activity concentration of the stock solution was 3.75 \replaced{MBq}{kBq}/ml, for a total activity of 604 MBq. However, gamma counter samples taken from the different spheres after the SPECT measurements showed a slightly lower activity concentration than those taken from the stock solution (mean decrease 1\%, max 2\%). This is likely caused by a slight dilution of the activity due to small amounts of water left in the spheres before filling, but a small amount of adsorption to the walls cannot be ruled out. This difference was added to the uncertainty budget, resulting in a total uncertainty of 4.0\% (k=2). 

\subsection{Radionuclide Calibrators}
In total, data were reported for eight clinical RNCs at seven different sites, including five VIK-202 chambers and three CRC-55tR chambers (Table \ref{tab:RNCs}). No data were available for site C, and site D only performed measurements using the clinical protocol. All RNCs were up to date with the quality controls required by Belgian regulations, including constancy and linearity tests, and regular verification of the accuracy using a traceable reference source. All sites used a high-energy \cs\, reference source, only some sites performed additional checks with a lower energy \isotope[57]{Co} \, source. Additionally, the accuracy of each chamber was verified at least once using a \lu\, source, and all sites except A and B determined a new dial setting based on this measurement. However, all sites \replaced{obtained their \lu \, vial}{used a \lu\, vial obtained} from a radiopharmaceutical company, often in a non-standardized geometry and without a clear traceability chain and associated uncertainty, instead of a calibrated source from a reference laboratory. Therefore, the \lu\, settings of none of the clinical RNCs could be considered traceable to primary standards, and no meaningful uncertainty could be determined.

At most sites, the clinical measurement protocol only differed from the factory protocol by the use of a different dial setting. Exceptions were site B, where a 3D-printed insert was used to elevate the dipper such that the vial was measured in the most sensitive region of the chamber, and site D , where a geometry correction factor of 98\% was used for all vials. None of the sites used correction factors for different vial types or filling volumes.

\begin{table}
    \caption{Overview of the reference and clinical RNCs included in this study.
    }\label{tab:RNCs}
    \begin{tabular}{cccccc}
    \toprule
        \multirow{2}{*}{ID} & \multirow{2}{*}{Chamber type} & \multicolumn{3}{c}{Dial settings} & \multirow{2}{*}{Calibration method}   \\ \cmidrule{3-5}
        ~ & ~ & Factory & Clinical & Corrected & ~ \\ 
    \midrule
        Ref 1 & VIK-202 &  751 & 763 & - &  Fidelis  \\ 
        Ref 2 & CRC-55tR  & 456 & 449 & - & Fidelis  \\ 
        A & VIK-202 & 751 & 751 & 758 & ITM CV  \\
        B & VIK-202 & 751 & 751 &  764/769\footnotemark[1] &  ITM CV  \\ 
        D & VIK-202 & - & 765 & 774 & Novartis CV  \\
        E & VIK-202 & 751 & 761 & 765 & Isotopia RV  \\ 
        F & CRC-55tR & 456 & 460/435\footnotemark[2] & 463 & Novartis CV/RP CV\footnotemark[2]\\
        G & VIK-202 & 751 & 774 & 768 & Novartis RV  \\ 
        H1 & VIK-202 & 751 & 762 & 752 & Novartis RV  \\ 
        H2 & CRC-55tR & 450 & 456 & 436 & Novartis RV \\ 
    \bottomrule
    \end{tabular}
    \footnotetext{\added{The factory dial settings were pre-programmed by the manufacturer, but most sites used a different clinical setting after calibration with a clinical vial (CV) or reference vial (RV) from a radiopharmaceutical company. The clinical dial settings for Ref1 and Ref2 were obtained through calibration against the Fidelis secondary standard. A corrected dial setting was calculated based on the measurements in this study.} A multiplication factor of 10 was applied on all RNCs. \deleted{CV: clinical vial, RV: reference vial}}
    \footnotetext[1]{Site B used an elevated vial positioning in the clinical protocol, resulting in different corrected dial settings depending on the position.}
    \footnotetext[2]{Site F receives \lu-PSMA from Novartis and \lu-DOTATATE from the radiopharmacy \added{(RP)} at another hospital resulting in two different clinical dial settings.}
\end{table}

At the time of measurement, the reference activity ranged from 291 to 66 MBq and from 158 to 35 MBq for the reference and clinical vials respectively. All RNCs showed excellent reproducibility, with a standard deviation smaller than 0.4\% for both vials and all RNCs, regardless of the measurement protocol. The measurement variation for the clinical vial was often smaller than for the reference vial, despite its lower activity, and it did not increase for the RNCs tested at the end of the measurement campaign when the activity in both vials was lower. This indicates that noise is limited for \lu \, in the activity range used in this study. 

\begin{figure}
    \centering
    \includegraphics[scale = 0.55]{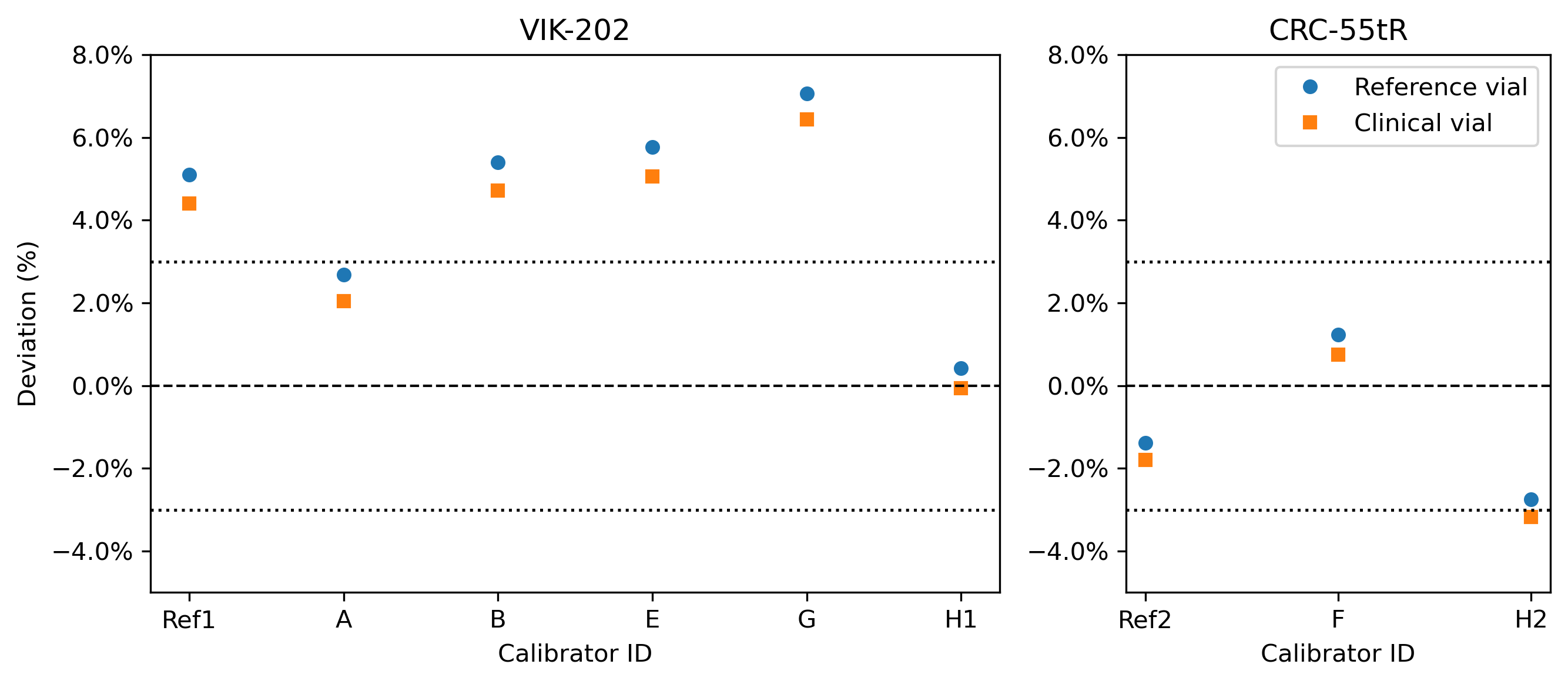}
    \caption{Deviations of the activities measured by the different RNCs using the factory protocol. The dotted lines indicate the 95\% confidence interval of the reference activity.}
    \label{fig:RNCs_factory}
\end{figure}

\begin{figure}
    \centering
    \includegraphics[scale = 0.55]{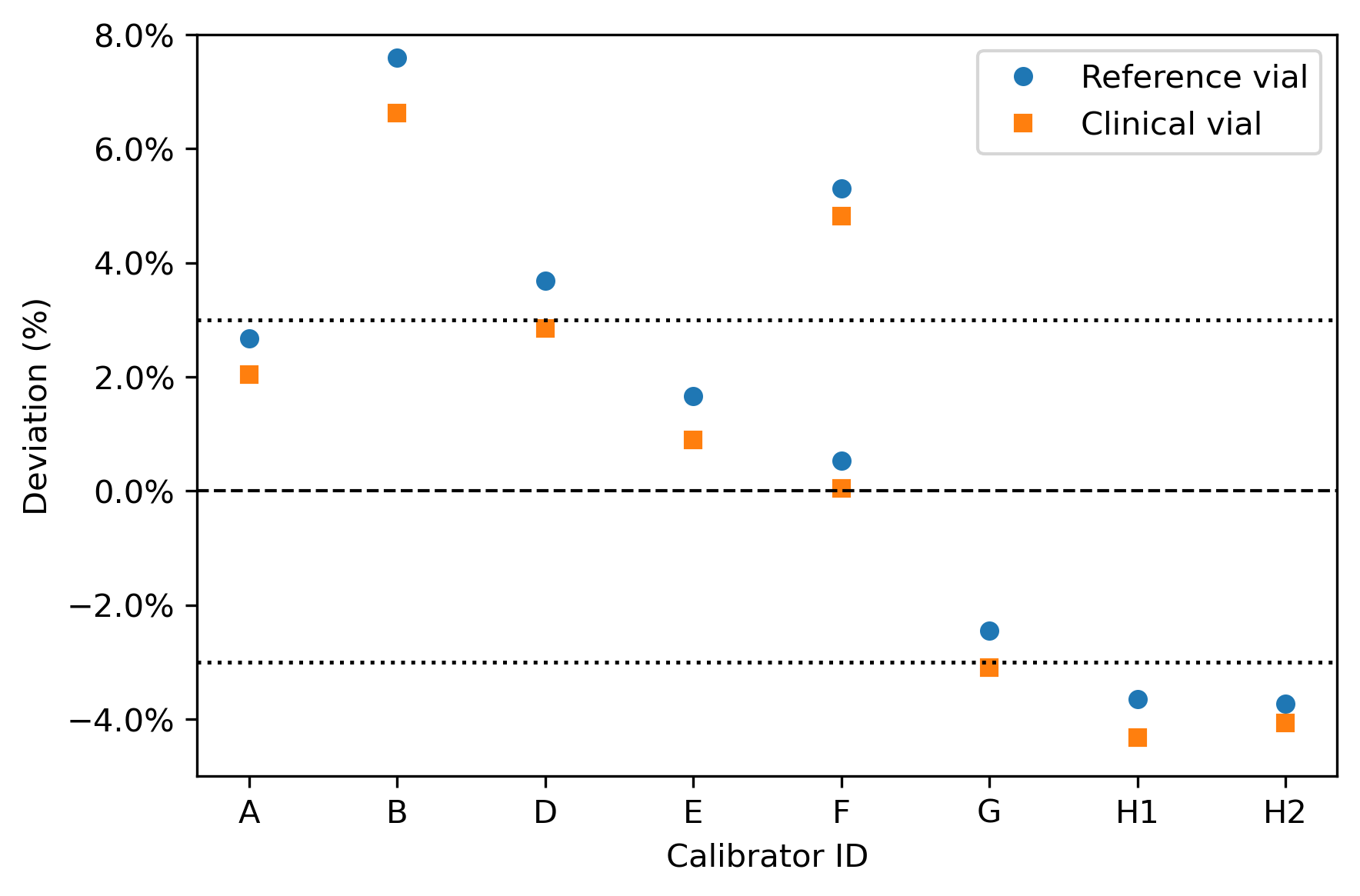}
    \caption{Deviations of the activities measured by the different RNCs using the clinical protocol. The dotted lines indicate the 95\% confidence interval of the reference activity. The upper and lower values for RNC F were obtained using the DOTATATE and PSMA settings respectively.}
    \label{fig:RNCs_clinical}
\end{figure}

The reported activities are compared to the reference activity in Figures \ref{fig:RNCs_factory} and \ref{fig:RNCs_clinical}. The factory protocol of the VIK-202 chambers overestimated the activity in the reference vial by $+0.4\%$ to $+7.1\%$, while for the CRC-55tR chambers the deviation from the reference activity ranged from $-3.7\%$ to $+1.2\%$. When using the site-specific clinical protocol, the activity deviations ranged from $-3.6\%$ to $+7.6\%$. The relative response to the clinical vials was always slightly lower than to the reference vial, with differences ranging from $-0.5\%$ to $-1.0\%$ and from $-0.3\%$ to $-0.5\%$ for the VIK-202 and CRC-55tR chambers respectively. The difference in response was similar for the clinical and factory protocols.

Table \ref{tab:RNCs} shows corrected dial settings that were calculated based on these deviations, following the instructions in the manuals of the two vendors and taking into account the different high gain settings for the VIK-202 chambers (see Appendix \ref{app:calibration_details}). The dial settings calculated from the measurements using the clinical and factory protocols were identical, except for site B, where, the elevated vial placement in the clinical protocol increased the response by about 2\%, leading to a slightly different corrected dial setting compared to the standard positioning. Similarly, geometry correction factors were estimated for each RNC, see Appendix \ref{app}. Results were consistent with the geometry correction factors for the reference RNCs. In particular, all correction factors were smaller than $1\%$.

\subsection{SPECT/CT}
In total, 13 SPECT/CT scanners (9 conventional and 4 CZT systems) were evaluated. All systems were up to date with the quality controls required by the manufacturer and Belgian regulations. Clinical protocols were available on 12 of the included systems, as summarized in Table \ref{tab:SPECT_systems}. All CZT systems used the acquisition protocol recommended by the manufacturer, while conventional systems used a range of different acquisition protocols, with 3 systems already using the standardized protocol. Clinical reconstruction settings differed from the standardized protocol in all cases. \added{Representative slices for the images acquired using the different scanners and different protocols are shown in appendix \ref{app}}

\begin{sidewaystable}
    \caption{Clinical acquisition and reconstruction protocols for each scanner.}
\label{tab:SPECT_systems}
    \begin{tabular}{cccccccccccc}
    \toprule
        \multicolumn{3}{c}{Scanner} & \multicolumn{2}{c}{Acquisition} & \added{\multirow{2}{*}{Matrix}} & \multicolumn{4}{c}{Reconstruction} & \multirow{2}{*}{Quantification} & \added{Software}\\
        \cmidrule(rl){1-3} \cmidrule(rl){4-5} \cmidrule(rl){7-10} 
        ID & Type & Crystal & Views & SC &  & Type & Algorithm & Updates & Post filter & ~ & \added{version}\\
    \midrule
        A1 & Symbia Intevo Bold & 3/8" & $2\times 60$ & TEW & 256m & Vendor & OSCGM & 16i5s & Gauss 8mm & xSPECT & \added{VB21A}\\
        A2 & Symbia T16        & 3/8" & $2\times 60$ & TEW & 256m & Vendor & OSEM & 16i5s & Gauss 8mm  & None & \added{VB10E}\\ 
        B1 & Discovery 870     & 5/8" & $2\times 30$ & DEW & 128m & Vendor & OSEM & 6i10s & None & ICF & \added{Smartconsole 1.0}\\ 
        B2 & Starguide         & 7.25mm & NA         & DEW & 260m & Vendor & BSREM+RDP & 20i10s & None & ICF & \added{Smartconsole 1.0}\\ 
        C1 & Symbia T16        & 3/8" & $2\times 32$ & TEW & 128m & Vendor & OSEM & 16i16s & None & ICF & \added{VA60E}\\ 
        C2 & Veriton \added{200}          & 6mm & NA           & TEW\footnotemark[3] & 256m & Vendor & OSEM & 6i8s & Median 3 & CF & \added{2.5} \\
        D  & Discovery 670     & 3/8" & $2\times 60$ & TEW & 256m & Vendor & OSEM & 2i10s & Butterworth \added{0.48/10} & None & \added{Xeleris 4}\\ 
        E1 & Discovery 670     & 3/8" & $2\times 30$ & DEW & 128m & MIM & OSEM & 12i10s & None & \replaced{MIM}{Spectra} & \added{7.3.2 / Xeleris 3}\footnotemark[3]\\ 
        E2\footnotemark[1] & Discovery 670Pro & 3/8" & - & - & - & - & - & - & - & None & \added{Xeleris 3}\\ 
        \multirow{2}{*}{F\footnotemark[2]} & \multirow{2}{*}{Symbia Intevo Bold} & \multirow{2}{*}{3/8"} & \multirow{2}{*}{$2\times 64$} & \multirow{2}{*}{TEW} & \multirow{2}{*}{128m} & MIM & OSEM & 8i16s & None & MIM & \added{7.2.3}\\ 
        ~ & ~ & ~ & ~ & ~ & ~ & Vendor & OSEM & 70i1s & Gauss 20.8mm & None & \added{VB21B}\\ 
        G & Starguide          & 7.25mm & NA         & DEW & 210m & Vendor & BSREM+RDP & 20i10s & None & ICF & \added{SmartConsole 1.0}\\
        H1 & Symbia Intevo Bold & 3/8" & $2\times 64$ & TEW & 128m & Vendor & OSEM & 6i8s & Gauss 8.4mm & None & \added{VB22A}\\ 
        H2 & Veriton  \added{200} & 6mm & NA           & TEW\footnotemark[3] & 256m & Vendor & OSEM & 4i8s & Median 3 & None & \added{2.5}\\
    \bottomrule
    \end{tabular}

    \footnotetext{The Veriton systems could not image the 208 keV peak because of an energy range cutoff at 200 keV. Therefore, the 113 keV peak was used instead. \added{Starguide systems used two bed positions, all others used one. For all conventional systems, the size of the acquisition and reconstruction matrix were indentical. For CZT systems, the matrix size refers to the reconstructed images only, as the pixel size during acquisition is fixed by the detector hardware. Similarly, the number of views (including swivel motion) was automatically determined by the system based on the phantom geometry.} OSCGM: Ordered Subset Conjugate Gradient Maximization, OSEM: Ordered subset Expectation Maximization, BSREM+RDP: Block Sequential Regularized Expectation Maximization with a Relative Difference Prior.}
    \footnotetext[1]{System E2 was not used clinically for \lu, so only the standardized protocol was used.}
    \footnotetext[2]{For scanner F, two clinical reconstruction protocols were available, with vendor reconstructions being used for clinical review by physicians, while MIM reconstructions were used for dosimetry.}
    \footnotetext[3]{\added{MIM 7.3.2 was used for the non-standardized recons, while Xeleris 3 was used for the (semi-)standardized recons}}
\end{sidewaystable}

\begin{figure}
    \centering
    \includegraphics[scale = 0.55]{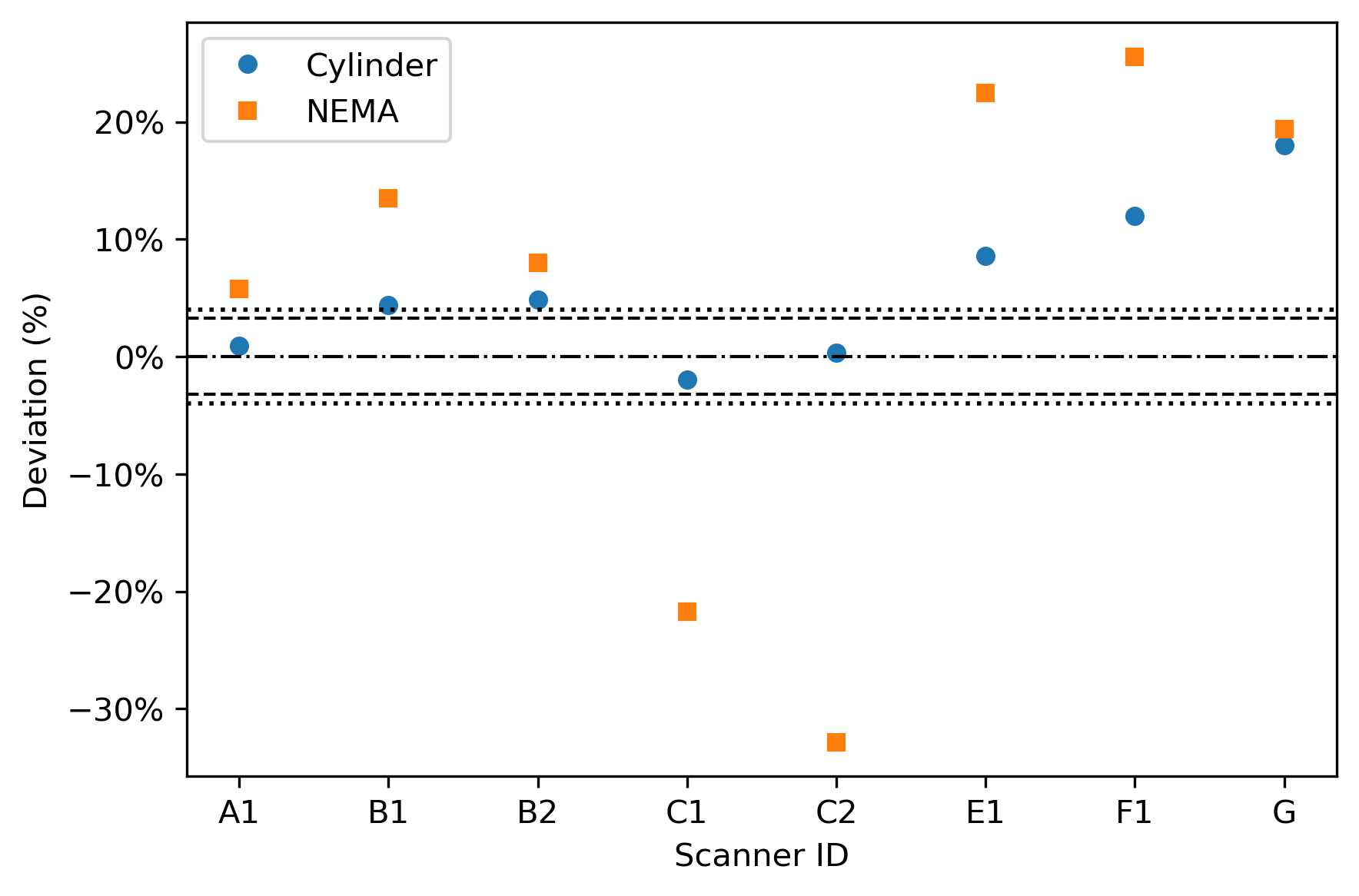}
    \caption{Deviations of the activities reported by the sites, using their own SPECT calibration factors, from the activity obtained with the gamma counter for 8 SPECT/CT systems. The dashed and dotted lines indicate the 95\% confidence intervals for the reference activity in the cylinder and NEMA phantom, respectively.}
    \label{fig:SPECT_quant}
\end{figure}

Image-based activity estimates for the two phantoms were reported for 8 scanners, as shown in Figure \ref{fig:SPECT_quant}. In three cases, sites reported measuring the activities directly from quantitative images in Bq/ml that were available through the Siemens xSPECT quant or MIM Spectra quant (MIM software) procedures. \added{By default, both Veriton systems produced images in Bq/ml as well. However, this quantification was off by a constant factor, so a previously determined CF was still needed to properly quantify the images. At the time of measurement, an up to date CF was only available for one of the two Veritons.} For the \replaced{remaining}{other} systems, sites \replaced{quantified}{had to scale} the images \deleted{by} using their own previously determined ICF \deleted{or CF}. Quantification errors ranged from -2\% to +18\% for the cylinder and from -33\% to +26\% for the NEMA phantom.

\begin{figure}
    \centering
    \includegraphics[scale= 0.55]{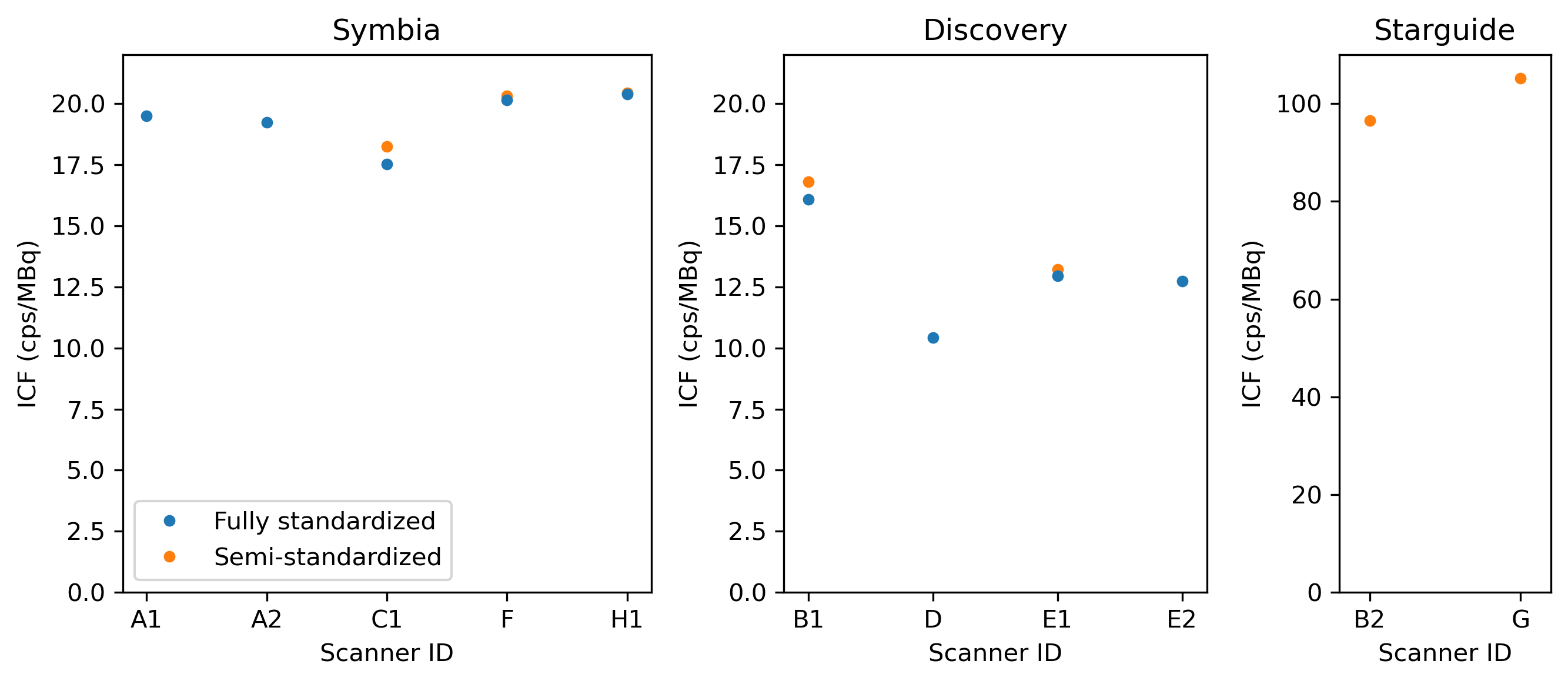}
    \caption{Image calibration factors (ICFs) for the conventional Siemens Symbia and GE Discovery systems and the CZT-based GE StarGuide system.}
    \label{fig:ICFs}
\end{figure}

The ICFs determined during this study are shown in Figure \ref{fig:ICFs}. Some Discovery reconstructions apply a scale factor of 4 to the projections before reconstruction when using resolution recovery.\footnote{This factor can be turned off using the setting "Projections multiplication" on the console. If no scale factor is applied, the Image Comments field in the DICOM header contains "MultProj 0".} For these images, the ICFs were divided by 4 for comparison with other systems. Only ICFs for the fully and semi-standardized scenarios are shown, since the choice of reconstruction protocol did not significantly affect the results. Only smoothing the scatter windows before reconstruction slightly reduced the ICF by increasing the scatter estimate, as discussed in appendix \ref{app:scatter_window_smoothing}. \added{Since the Veriton systems directly produced images in Bq/ml, no ICF could be determined.}

\deleted{For the Veriton systems, all reconstructed images \replaced{had voxel values in}{were proportional to} Bq/ml instead of counts, so no ICF could be determined. \replaced{However, a}{The global} CF was \added{still} needed to properly quantify the images\added{. This factor} depended strongly on the local system calibration, software version and protocol choice.} 

\begin{figure}
    \centering
    \includegraphics[width = \linewidth]{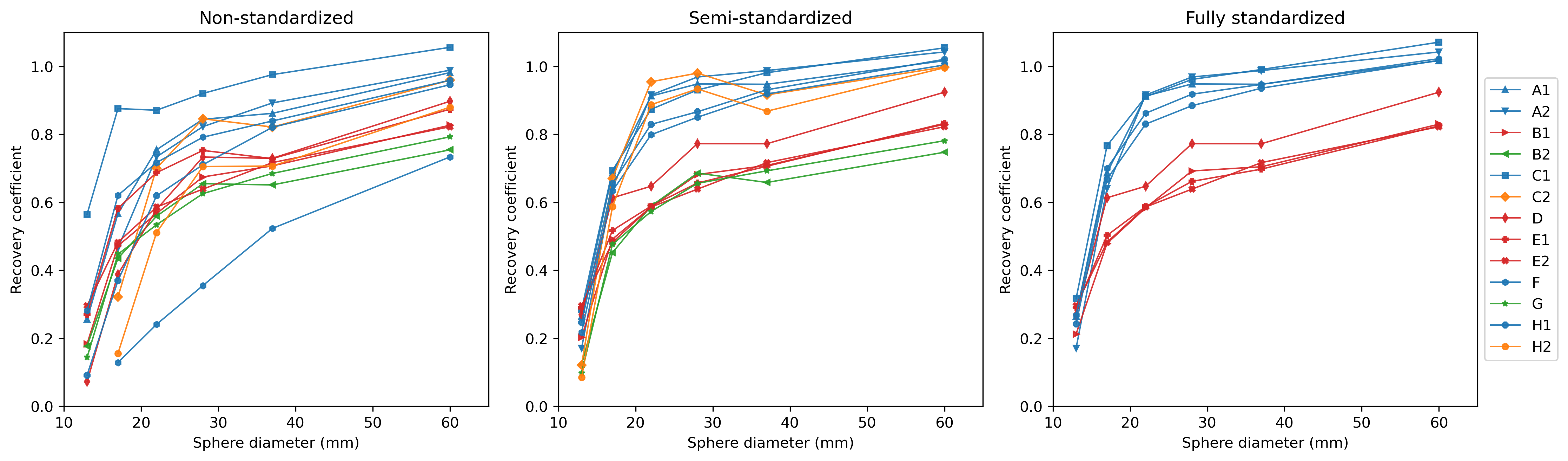}
    \caption{
Recovery curves for the different systems using the different protocols. Blue: Siemens Symbia (n=5); red: GE Discovery (n=4); purple: GE Starguide (n=2); orange: Spectrum Dynamics Veriton (n=2). Since the standardized acquisition protocol could only be applied to conventional systems, no fully standardized curves for the CZT systems are shown.}
    \label{fig:RCs_overview}
\end{figure}

Recovery curves for the different systems and levels of standardization are summarized in Figure \ref{fig:RCs_overview}. While the overall RC range is very wide for all levels of standardization, separate bands per system family (Siemens Symbia, GE Discovery, GE Starguide, and Spectrum Dynamics Veriton) appear for the semi-standardized images, which are shown in Figure \ref{fig:RC_bands}. Numerical values and recovery bands for the fully standardized images are shown in Appendix \ref{app}. Figure \ref{fig:SSvsFSx2} shows a side-by-side comparison of the fully and semi-standardized RCs for \added{two of} the five conventional systems whose clinical acquisition protocol differed from the standardized protocol\added{, curves for the other three are shown in Figure \ref{fig:SSvsFSx3}}.

\begin{figure}
    \centering
    \includegraphics[width=0.8\linewidth]{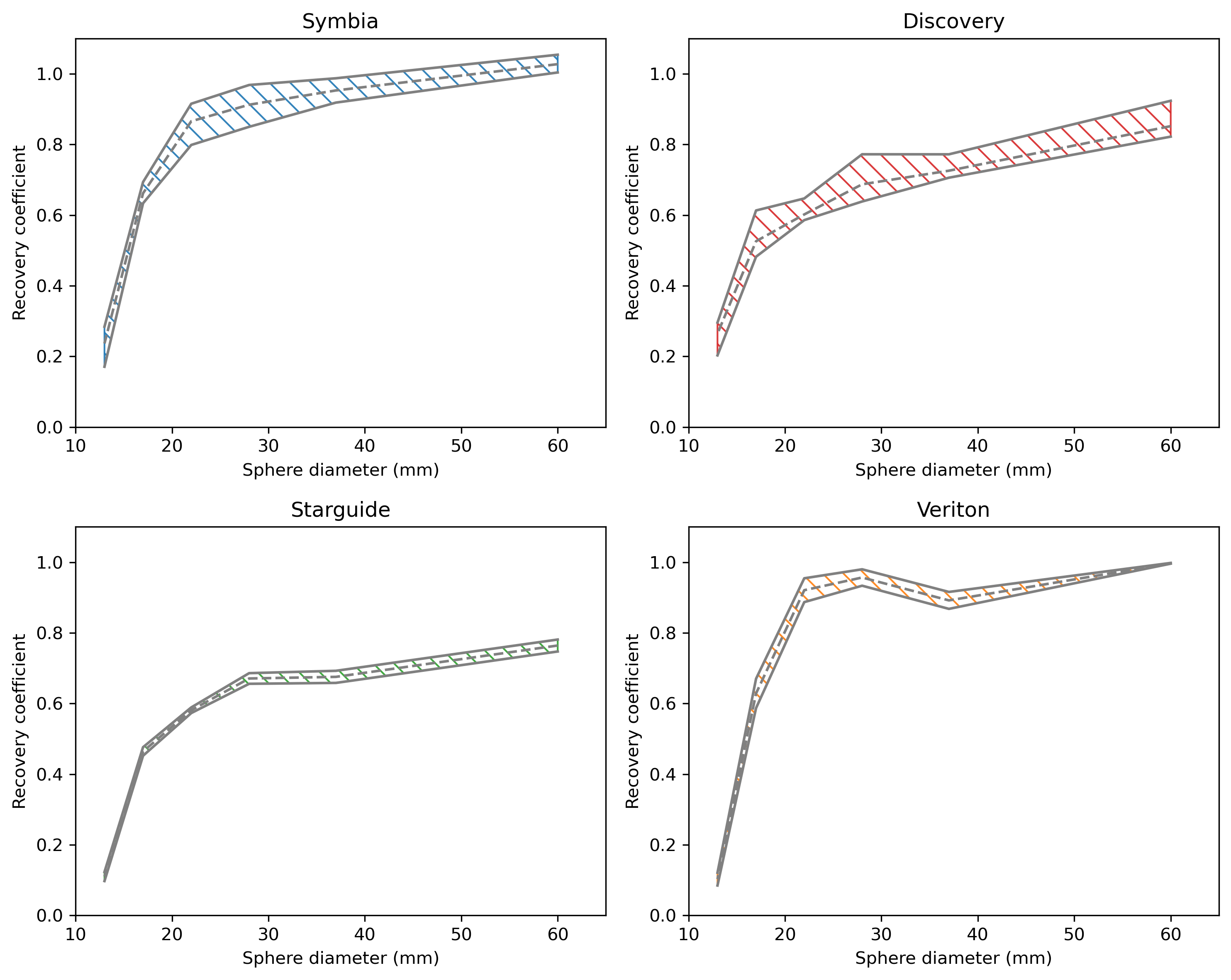}
    \caption{Recovery coefficient (RC) bands for the semi-standardized SPECT images (clinical acquisition with standardized reconstruction) of the different scanner families. The dashed line denotes the mean RC, and the outer lines indicate the RC range.}
    \label{fig:RC_bands}
\end{figure}

\begin{figure}
    \centering
    \includegraphics[width=0.8\linewidth]{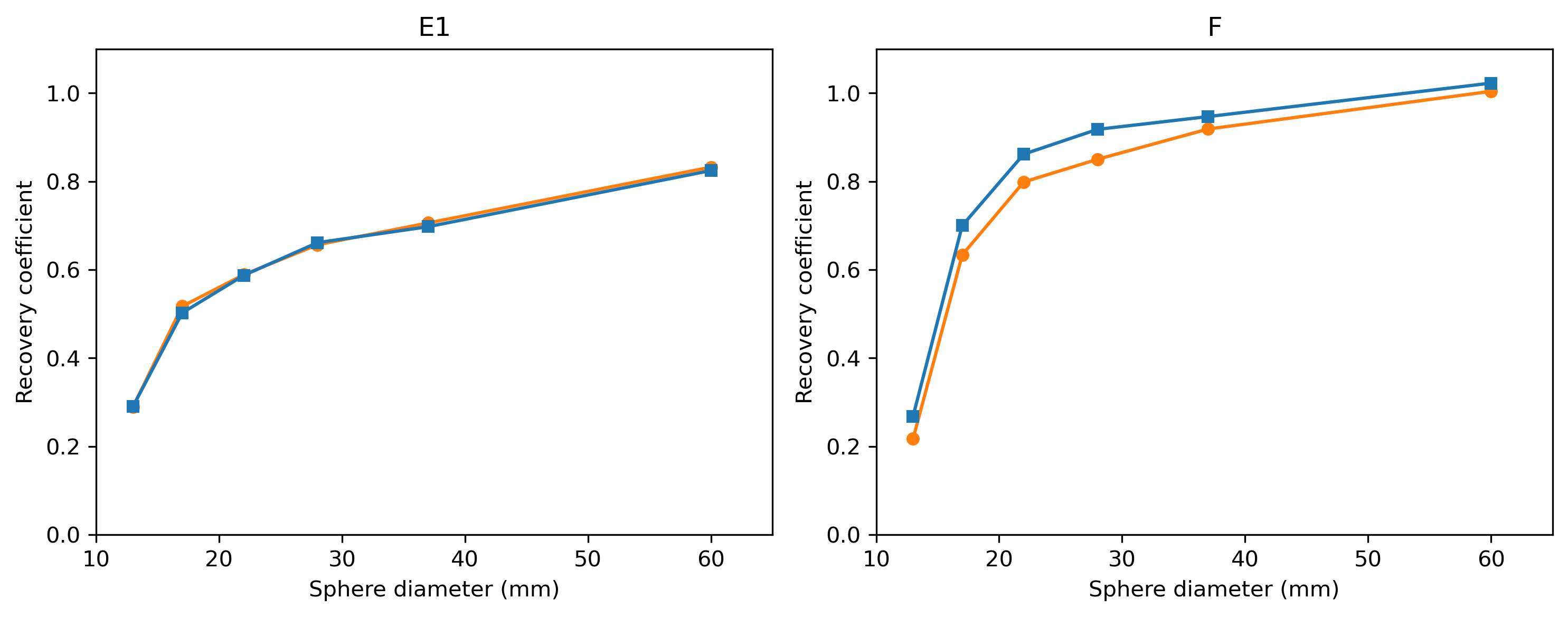}
    \caption{Side-by-side comparison of the recovery curves \added{of systems E1 and F} using fully standardized (blue) versus semi-standardized protocols (orange). \replaced{Equivalent curves for systems B1, C1 and H1 are shown in appendix \ref{app}.}{Curves for the different scanners are shifted in the y direction for clarity.}}
    \label{fig:SSvsFSx2}
\end{figure}

\begin{figure}
    \centering
    \includegraphics[width=0.8\linewidth]{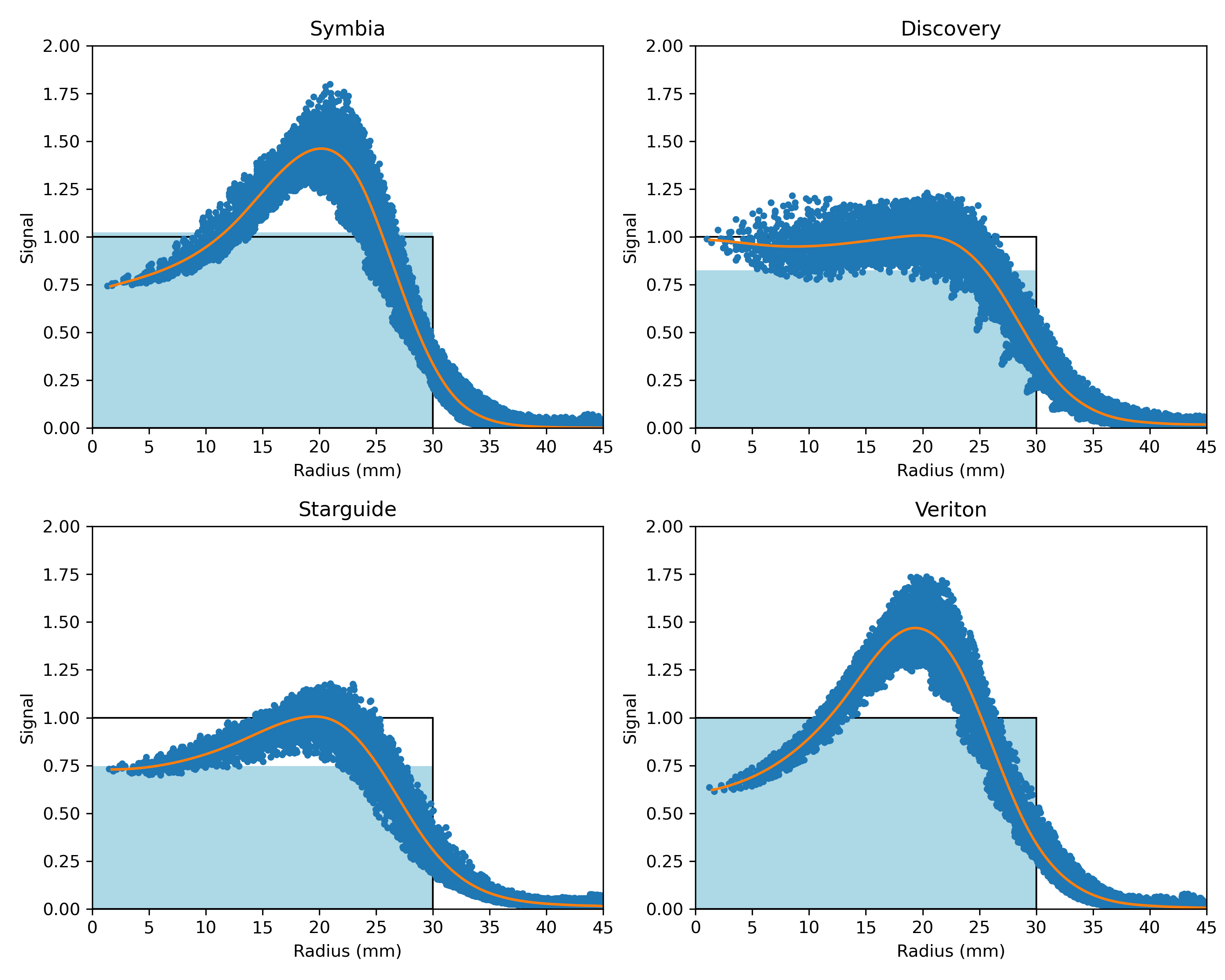}
    \caption{Gibbs artifacts in the largest sphere for the different system types in the fully standardized scenario. Representative examples are shown for each system type: Symbia - F, Discovery - E1, Starguide - B2, Veriton - H2). Each dot represents one voxel, and a smoothing spline is fitted through the data. The profiles are normalized such that the true signal is 1 inside the sphere and 0 outside. The light blue background indicates the mean signal over the sphere volume, corresponding to the recovery coefficient.}
    \label{fig:GA_comparison}
\end{figure}

Radial profiles and Gibbs artifacts in the largest sphere are shown in Figure \ref{fig:GA_comparison} for representative examples of each system type. Numerical values for the Gibbs artifacts are given in Table \ref{tab:GA_strength}. No values for GE Discovery images are shown, since none of them showed significant Gibbs artifacts (GA$<5\%$).

Representative image slices of the cylindrical and NEMA phantoms acquired using the non-standardized and fully standardized protocols are shown in Appendix \ref{app}.

\begin{table}
\centering
\caption{Strength of the Gibbs artifact (GA) in the largest sphere of the NEMA IQ phantom for the different SPECT systems, shown for the non-standardized (clinical acquisition and reconstruction), semi-standardized (clinical acquisition and standardized reconstruction), and fully standardized (standardized acquisition and reconstruction) SPECT imaging scenarios.}\label{tab:GA_strength}
\begin{tabular}{c c c c c}
\toprule
Type & ID & Non-standardized & Semi-standardized & Fully standardized \\
\midrule 
\multirow{5}{*}{Siemens}
    & A1 & 43.4 &    -    & 34.6 \\
    & A2 & 17.0 &    -    & 25.2 \\
    & H1 & 17.7 & 33.9 & 32.8 \\
    & C1 & 30.6 & 39.2 & 40.0 \\
    & F  & 18.9\footnotemark[1] & 33.8 & 32.7 \\
\multirow{2}{*}{Starguide}
    & B2 & 14.6 & 16.0 & -  \\
    & G  & 12.7 & 17.2 & - \\
\multirow{2}{*}{Veriton}
    & C2 & 22.3 & 35.5 & - \\
    & H2 & 9.7  & 40.5 & - \\
\bottomrule
\end{tabular}
\footnotetext{The GE Discovery systems were excluded, since none of them showed clear Gibbs artifacts.}
\footnotetext[1]{GA strength for the clinical MIM reconstruction. The clinical vendor reconstruction did not show any Gibbs artifacts.}
\end{table}

\section{Discussion}
\subsection{Radionuclide calibrators}
Two vials with traceable activities were used to evaluate 8 clinical RNCs at 7 different hospitals. All sites reported adherence to the quality assurance procedures required by Belgian regulation, but since this regulation only requires testing using long-lived and diagnostic radionuclides,  \lu-specific procedures were minimal and differed substantially across hospitals. In particular, none of the clinical RNCs had \lu\, calibrations traceable to a primary standard, resulting in an $11\%$ spread in activity measurements using the clinical protocols (Figure \ref{fig:RNCs_clinical}). 

A similar spread in activities was observed when using the factory protocols (Figure \ref{fig:RNCs_factory}). While part of this was due to VIK-202 chambers overestimating and CRC-55tR chambers underestimating the activity on average, differences were significant even between RNCs of the same type. This suggests that small differences between nominally identical chambers, for example due to manufacturing (e.g. variations in wall thickness), installation (e.g. presence of lead shielding), or drift over time, result in non-negligible differences in activity readings. Therefore, dial settings obtained for one chamber cannot simply be transferred to other chambers of the same type (Table \ref{tab:RNCs}).

The observed range in activity measurements could be considered acceptable when determining the injected activity, but when using RNCs to calibrate quantitative SPECT equipment, it is only one of the many factors affecting the uncertainty in the final result. Since the reference activity in the vials could be determined to within 3\% through careful calibration with respect to a secondary standard, the variation in activity measurements (and \added{absorbed} dose estimates) could be significantly reduced by requiring stricter calibration procedures for all clinical RNCs. In particular, calibration using activity supplied by radiopharmaceutical companies or using standardized dial settings should be avoided\replaced{ in favor of}{, and} direct calibration to a secondary standard such as the Fidelis\deleted{ should still be regarded as the gold standard}. Alternatively, traceable reference sources like the vials used in this study could be measured across multiple hospitals to harmonize RNC calibration between centres. \added{However, this alternative approach has a longer calibration chain which increases the associated uncertainties, so direct calibration to a secondary standard should remain the preferred method.}

The results for the clinical vial were very similar to those for the reference vial with differences in response due to geometry always below 1\%, showing that for \lu \, measurements, the overall calibration of the chamber is more important than accounting for different vial geometries. This is because the relevant photon energies of \lu \, are relatively high, so attenuation within the vial is limited for all geometries. For isotopes with lower energy photons like \tb, geometry effects have been shown to be more important, requiring significant corrections for different vial types \cite{juget}.

\added{A previous study by Saldararriaga Vargas et al. investigated RNC accuracy for several radionuclides including \lu \, in hospitals in the Netherlands, Germany and Belgium with respect to 5\% and 10\% suspension levels \cite{Clarita2}. They found that a quarter of \lu-measurements deviated more than 5\% from the reference value, while none deviated more than 10\%. This is in line with our results, despite most of their measurements using the factory dial settings. This confirms that calibration using vials from radiopharmaceutical companies does not improve average RNC accuracy across multiple centers. Furthermore, differences between vials and syringes were found to be small, confirming that errors due to geometry effects smaller than typical calibration errors for \lu.}

\subsection{SPECT/CT}
A uniform cylinder and a NEMA phantom filled with \lu \, were used to evaluate 13 SPECT/CT scanners at 8 different sites, including 4 3D CZT systems. \replaced{Special care was taken to ensure the activities in both phantoms were traceable to secondary standards, similarly to previous work by Robinson et al . \cite{Robinson}. However, they implemented the RNC-based phantom filling procedures described in \cite{gpg2}, while our method relied on a calibrated gamma counter. Besides achieving similar confidence levels, our approach also allowed to verify that the activity in the phantoms remained uniform throughout the two-week measurement campaign.}{Through measurements with a calibrated gamma counter, the activity in both phantoms could be accurately determined and it could be verified that the activity remained uniform throughout the two-week measurement campaign.}

Similar to the RNCs, all sites reported comparable quality assurance programs based on Belgian regulations. However, specific tests with \lu \, were limited and non-standardized, as they are not explicitly required by the current regulations. Furthermore, not all scanners were calibrated for \lu, since quantitative SPECT and dosimetry are not yet part of routine clinical practice. Therefore, quantitative measurements were only reported for 8 out of 13 scanners (Figure \ref{fig:SPECT_quant}).

\replaced{The overall system calibration of these 8 scanners was evaluated using the cylindrical phantom.}{For the cylindrical phantom,} \replaced{Deviations in the reported activity}{deviations} were below 5\% for 5 out of 8 scanners, which is of the same order as the observed variation in the RNC measurements. This is expected, since most systems were calibrated with a uniform cylinder phantom with an activity determined using the in-house RNC. Only the calibration system A1 did not rely on a RNC, since the the \lu \, sensitivity was indirectly derived from a traceable \isotope[75]{Se} reference source (Eckert \& Ziegler) using the xSPECT Quant method. \added{Attempting to correct the quantification results for the RNC deviations observed in this study did indeed improve the accuracy for all but one scanner, as shown in Figure \ref{fig:Spect_Quant_corrected}}. \replaced{Still,}{For three other scanners,} the quantification errors \added{of three scanners} could not be \added{fully} explained by the RNC measurements alone, indicating that either the calibration procedure or the quantification methods introduced additional errors.

\replaced{A similar exercise was performed using the NEMA phantom, yielding deviations}{Quantification errors for the NEMA phantom were} larger than \deleted{those} for the cylindrical phantom in all cases, and large variability between sites was observed. Here, the small sphere sizes cause significant spill-out, which was handled in different ways by the different sites. Site C did not account for spill-out, resulting in an underestimation of the total activity. All other sites used some form of expanded VOI, ranging from expanding the spheres by a fixed margin to including all activity within the entire image. While these approaches do account for spill-out, they can also include uncorrected scatter or residual background activity due to septal penetration. \replaced{This shows that quantification in complex geometries does not only rely on overall scanner calibration, but is also highly dependent on calculation methods such as partial-volume correction. A full comparison of post-processing methodologies was outside the scope of this study, but previous}{Previous} research found similar variations in quantification when using \replaced{site-specific processing}{clinical} protocols \cite{Wevrett}, while others have shown that \replaced{quantification performance in complex geometries}{performance} can be harmonized by using a standardized methodology \cite{MRT}. \deleted{This highlights that quantitative accuracy is not only determined by scanner calibration, but also by site-specific calculation methods \cite{Ramirez}.} Therefore, harmonization of dosimetry calculations \replaced{is also}{might also be} required for \added{absorbed} dose estimates that are fully comparable between different hospitals.

Centralized analysis of the cylinder data showed that reconstruction settings had no impact on the ICF, except for scatter window smoothing (Appendix \ref{app:scatter_window_smoothing}). Similarly, changing the matrix size or number of views in the acquisition protocol had a negligible impact on the ICFs (Figure \ref{fig:ICFs}). The only important factor was the choice of energy windows. In particular, \added{dual energy window (}DEW\added{)} scatter correction yielded slightly higher ICFs than \added{triple energy window (}TEW\added{)}, since excluding the upper window decreases the number of counts in the scatter estimate, resulting in a higher number of counts in the reconstructed image. Still, this effect was rather small, ranging from 2.2\% to 4.6\% for scanners E1 and B1, respectively. 

ICFs were very similar for systems of the same type, apart from system B1 due to its thicker crystal. For the conventional SPECT/CT systems, ICFs were in line with the planar sensitivity of both heads combined derived from previous quality control measurements, so they can be interpreted as the system sensitivity in tomographic mode. For the Starguide systems, ICFs were also close to the combined planar sensitivity of all detectors. However, these detectors require a swivel motion to compensate for the limited field of view, so not all activity is seen by all detectors at all times. This leads to a lower effective sensitivity that depends on object size and acquisition mode, so the ICFs of the CZT systems cannot be directly compared to those of conventional systems.

Among the conventional systems, Symbia scanners had a significantly higher ICF than Discovery systems  ($\approx$ 20 cps/MBq vs $\approx$ 12 cps/MBq), which is due to differences in collimator design. As shown in Table \ref{tab:collims}, the GE collimator has longer holes, resulting in better geometric resolution at the cost of lower sensitivity. This was reflected in the increased acquisition time needed for Discovery systems to reach the same counts target as Symbia systems, and in the finer noise pattern in the GE reconstructions, resulting in a higher standard deviation in the central region of the cylinder (around 23\% and 13\%, respectively). Using a thicker crystal increased the ICF by about 25\%, in line with the calculated increase in \replaced{interaction probability}{stopping power}. These ICFs are in line with previously reported values \cite{MRT,SteffieP}.

\begin{table}
    \caption{Specifications of the medium-energy collimators used on the Discovery and Symbia SPECT/CT systems.}\label{tab:collims}
    \begin{tabular}{cccccc}
    \toprule
        ~& \shortstack{Hole diameter \\(mm)} & \shortstack{Septal thickness\\ (mm)} & \shortstack{Hole length\\ (mm)} & \shortstack{Sensitivity \\(\%)} & \shortstack{Resolution\\(mm)}  \\
    \midrule
        Discovery (MEGP) & 3 & 1.05 & 58 & 0.011 & 8.5 \\
        Symbia (MELP) & 2.94 & 1.14 & 40.64 & 0.021 & 10.7 \\
    \bottomrule
    \end{tabular}
    \footnotetext{Collimator parameters as provided by the manufacturer, with geometric sensitivity and resolution at 10 cm calculated for 208 keV following \cite{collim}}
\end{table}

The NEMA phantom was used to evaluate the image quality of the different systems using different acquisition and reconstruction protocols. Images generated using non-standardized protocols showed substantial visual variation, which translated into very different recovery curves (Figure \ref{fig:RCs_overview}) and Gibbs artifact strength (Table \ref{tab:GA_strength}). Differences in the number of iterations and subsets, post-filtering, and reconstruction type all contributed to this variation in RC curves. Consequently, applying a standardized reconstruction protocol in the semi-standardized scenario, markedly reduced this variability. However, standardizing acquisition protocol as well in the fully standardized scenario did not significantly reduce the remaining variability.

Comparing the fully and semi-standardized RCs for the five conventional SPECT systems whose clinical acquisition protocol differed from the standardized acquisition protocol (Figure \ref{fig:SSvsFSx2}) showed that changes in the number of views or the energy window settings have little impact on the reconstructed image quality. Only the matrix size appeared to play a small role for the Symbia systems, with systems C1 and F showing smaller RCs for the smaller spheres when using a 128×128 matrix compared to a 256×256 matrix (system H1 used a 128×128 matrix in both protocols). This is likely due to the fact that smaller voxels allow a better representation of sharp edges, which is more important for smaller objects. Since the radial profiles of the Symbia reconstructions are sharper at the sphere edges, this effect is more pronounced than for the GE reconstructions.

With standardized reconstruction settings, the RCs for each system type formed narrow bands (Figure \ref{fig:RC_bands}), and Gibbs artifacts were \replaced{comparable}{of the same order of magnitude}. This shows that small variations in manufacturing, maintenance, and electronics, have only minor impact on the resulting images as long as the main hardware and software design remains the same. Notably, system B1 falls within the same RC band as the other Discovery systems in this study despite having a 5/8" crystal, showing that the slightly poorer intrinsic resolution of a thicker crystal has negligible impact on images required with a medium energy collimator. As shown in appendix \ref{app:scatter_window_smoothing}, pre-filtering the scatter windows has a significant impact on the resulting images, so it remains important to \replaced{consider}{harmonize} all possible reconstruction settings, and not only the number of updates and post smoothing.

Even though the RC bands per system type are quite narrow, large differences can be observed between scanner types. First, the RCs for the Symbia systems were significantly higher than for the Discovery systems, even though the Discovery collimator has better geometric resolution than the Symbia collimator (Table \ref{tab:collims}). On the other hand, the Symbia images showed very strong Gibbs artifacts ($>30\%$), while those in the Discovery images were negligible. This suggests that there are differences in the implementation of the reconstruction algorithm, in particular resolution modeling, that strongly impact the reconstructed images. Using vendor-neutral software could help reduce this inter-system variability, but a recent study on \isotope[166]{Ho}-SPECT showed that even with identical reconstructions, differences between vendors persisted \cite{ho_intercomparison}. It is therefore expected that using vendor-neutral reconstructions with identical parameters cannot fully harmonize RCs and GAs in \lu-SPECT either .

Second, the very different design of the 3D CZT systems resulted in large differences compared to the conventional systems. One challenge is the fixed collimator, which is not optimized for \lu, resulting in significant septal penetration \added{(Figure \ref{fig:septal_penetration})} and decreased RCs when imaging the 208 keV peak using the Starguide system. The Veriton design avoids this problem by only imaging the 113 keV peak, but this increases the impact of scatter and scatter correction. \added{Furthermore, the overall shape of the RC curves was different compared to the conventional systems (Figure \ref{fig:RC_bands}). In particular, the mean recovery curves of the Symbia and Discovery systems increased with increasing sphere size, while the Starguide and Veriton systems showed RCs for the 28 mm sphere that were larger than or comparable to the 37 mm sphere. On the other hand, RCs in the smallest sphere were much smaller compared to the conventional systems. More generally, the CZT systems showed increased RCs in spheres at the top of the phantom, while the RCs in spheres at the bottom were decreased. It has previously been observed that RCs of conventional systems do not only depend on sphere size, but also on the full phantom geometry \cite{LeubeRC}. It is therefore likely that the RCs of CZT systems also vary with geometry, but that the exact dependence differs due to their different design. For example, the multiple detectors can better follow the contour of the top of the phantom than a conventional system, but are still limited by the bed for the bottom of the phantom. Furthermore, the swivel motion favor spheres close to the detector, since these can be seen from many different angles and are visible for longer for a fixed motion range.}

\deleted{Other differences include the fundamentally different geometry and the introduction of the swivel motion of the heads, which affects the depth-dependent resolution, leading to a different shape of the RC curves.}

\added{Finally, note that the Siemens and Veriton images achieve RCs close to or above 1 for the largest sphere. This indicates that the RCs are not purely determined by the effective resolution of the system, and that other effects, like sub-optimal performance of window-based scatter correction, could have an impact.}

Our data show that for \lu-SPECT/CT, it is possible to harmonize RC curves between systems of the same type by using \replaced{standardized}{harmonized} reconstruction settings, without requiring \replaced{standardized}{harmonized} acquisition protocols. Therefore, existing clinical protocols can still be used, as long as reconstructions with standardized settings are used when pooling data from different hospitals. Since the only acquisition setting found to have noticeable impact was matrix size, it could be worthwhile to always acquire data in a finer matrix (e.g. $256 \times 256$) and apply interpolation before reconstruction if needed. 

\replaced{Existing accreditation programs for PET/CT imaging \cite{EARLPET1,EARLPET2} go one step further and attempt to}{ The next step could be to} harmonize the RC curves of different system types as well\deleted{, similar to existing accreditation programs for PET/CT imaging} \deleted{\cite{EARLPET1}\cite{EARLPET2}}. These programs rely on image smoothing to bring the RCs into the required band, which is desirable for diagnostic images, since it ensures visually similar images across all system types. \added{However, the end goal of \lu-SPECT/CT harmonization should be reliable and comparable absorbed dose estimates across centers, which does not necessarily require visually consistent images. On the contrary, fully harmonizing RCs by post smoothing could be detrimental to absorbed dose estimates if it requires significantly }\deleted{However, this approach also requires} reducing  RCs for many scanners, lowering the effective resolution and increasing the partial-volume effect\deleted{, which is detrimental for images used in dosimetry}. Indeed, resolution is one of the main factors determining the uncertainty in \added{absorbed} dose calculations \cite{EANM_uncertainties,CZT2}, and correcting for the partial-volume effect remains challenging \cite{LeubePVC}. \added{Given the large differences between system types that we observed even after standardizing acquisition and reconstruction settings, it might be counterproductive to fully harmonize RCs across all different system types. More research into the effect of differences in RC curves on the uncertainties in absorbed dose estimates is needed, but it would} \deleted{Therefore, it would} be preferable to choose reconstruction settings that minimize the uncertainty in the resulting \added{absorbed} dose estimates for each scanner type, rather than aiming for visually homogeneous images across all systems. These settings would need to be determined in function of the specific quantification procedure and may differ between scanner types and between dosimetry applications.


\section{Conclusion}
This multicenter phantom study evaluated the quantitative accuracy and image quality for \lu\, of 9 conventional and 4 3D CZT systems at 8 different hospitals using both clinical and standardized protocols. While all sites had adequate \lu \, imaging capabilities, variations in absolute quantification were significant due to differences in calculation and calibration procedures, including non-traceable calibration of RNCs. Differences in image quality were large when using clinical protocols, but standardized reconstruction settings strongly reduced variability between systems of the same type, regardless of acquisition protocol. Differences between system types persisted when using standardized protocols due to differences in hardware and software design, most notably between conventional and CZT systems. Therefore, this research shows that it is possible to harmonize performance of systems of the same type by establishing \replaced{standardized}{harmonized} reconstruction protocols, but \replaced{this does not harmonize}{harmonizing} systems from different types \deleted{remains challenging}.

\backmatter

\bmhead{Supplementary information}




\bmhead{Acknowledgements}
We would like to thank Wies Deckers and Christelle Terwinghe for their support during the preparation and the execution of the measurements at UZ Leuven, and Matthijs Sevenois for the independent analysis of the UZ Leuven data. We would like to thank Cristian Mihailescu en Raf Aarts (LNK) for their contributions during the calibration of the reference RNCs with respect to the Fidelis secondary standard.

\section*{Declarations}

\begin{itemize}
\item Funding: This work was supported by the European Partnership for Radiation Protection Research (PianoForte) as part of EU’s “EURATOM” research and innovation program. Wies Claeys is supported by the Research Foundation Flanders (FWO) through grant 1S40025N.
\item Competing interests: The authors have no relevant financial or non-financial interests to disclose.
\item Ethics approval and consent to participate: Not applicable
\item Consent for publication: Not applicable
\item Data availability: The curated dataset of all SPECT/CT image data used in this study is available in the Zenodo repository (\url{https://doi.org/10.5281/zenodo.17897044}). All other data are available in the KU Leuven Research Data Repository (\url{https:/doi.org/10.48804/V5KQMX}).

\item Materials availability: Not applicable
\item Code availability: Python scripts used for image analysis are available in the KU Leuven Research Data Repository (ID to be added).
\item Author contribution: All authors performed measurements and provided technical expertise. Wies Claeys and Michel Koole designed the experiments and wrote the manuscript. Wies Claeys performed the data analysis. All authors read and approved the final manuscript.

\end{itemize}

\begin{appendices}
\section{Scatter window smoothing}\label{app:scatter_window_smoothing}
The initial standardized reconstruction settings did not include instructions on pre-filtering, resulting in different filters being applied in the reconstructions from different sites. In particular, Siemens images were reconstructed either without scatter filter or with 20 mm Gaussian smoothing, resulting in visual and quantitative differences between the images from different systems.

Previous research has shown that smoothing the scatter windows can improve the reconstructed image \cite{scatter_methods}\cite{scatter_smoothing}. The OSEM algorithm takes into account the Poisson noise in the sinogram data, so filtering of the photopeak window is generally not applied before reconstruction. The scatter windows on the other hand are not reconstructed directly, but used to get an estimate of the scatter which is added to the forward projection model (i.e. the denominator of the \replaced{OSEM}{MLEM} equation). This equation only holds for the expectation of the scatter, i.e. the scatter estimate should be noise-free. \replaced{Consequently, simulation studies have shown}{Therefore, literature suggests} that reducing the noise by filtering the scatter windows improves the accuracy of the reconstruction\added{\cite{scatter_methods,scatter_smoothing}}.

\begin{figure}
    \centering
    \includegraphics[width=\linewidth]{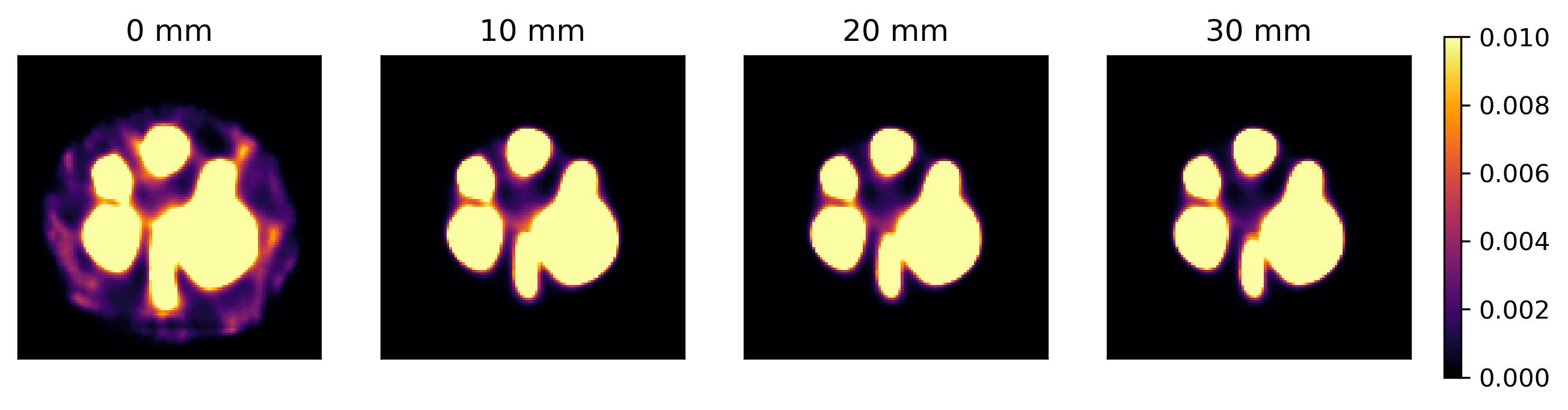}
    \caption{Reconstructions of the NEMA phantom with varying levels of Gaussian pre-filtering on the scatter windows. The data is rescaled such that the expected signal inside the spheres is 1.}
    \label{fig:SC_smoothing_NEMA}
\end{figure}

To  test the impact of pre-filtering the scatter windows in \lu\, reconstructions, the data from system A1 were reconstructed using the standardized reconstruction settings with a Gaussian pre-filter of increasing \added{full width at half maximum (}FWHM\added{)}. As shown in figure \ref{fig:SC_smoothing_NEMA}, the NEMA images without scatter smoothing appear to contain a small background signal, which is not present in the other reconstructions. These are likely scatter counts that are not properly compensated for, and contribute more than 5\% of the total counts in the image. This background signal disappears when applying a scatter filter, regardless of the FWHM. This indicates that the noise in the scatter windows leads to a bias in the \replaced{image}{scatter}, which can be reduced by pre-filtering the data.

\begin{figure}
    \centering
    \includegraphics[width=0.5\linewidth]{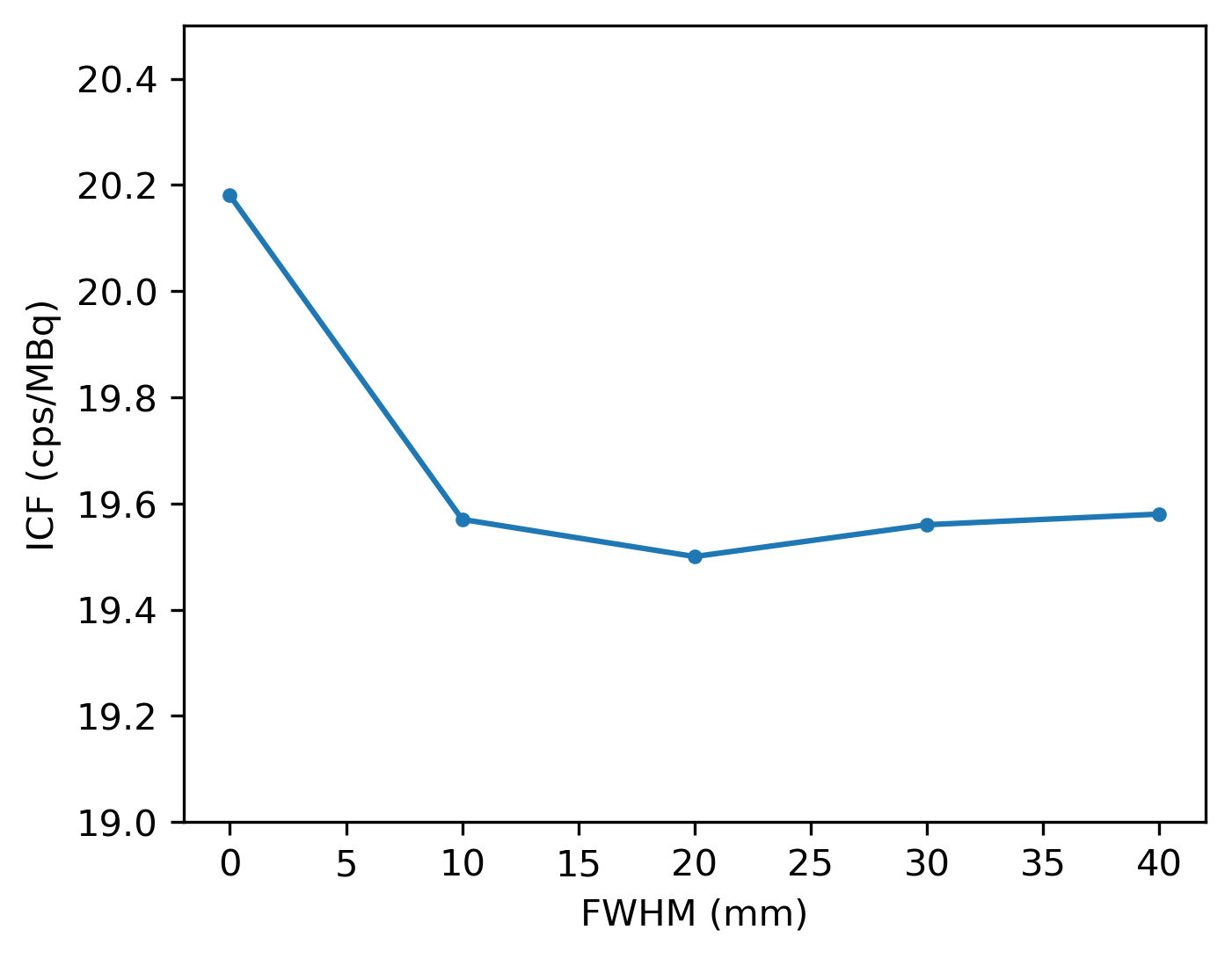}
    \caption{ICFs for different levels of scatter window smoothing.}
    \label{fig:SC_smoothing_ICFs}
\end{figure}

This is confirmed by the cylinder data: figure \ref{fig:SC_smoothing_ICFs} shows the ICFs obtained from the cylinder reconstructions using different pre-filter strengths. There is a noticeable drop in ICF when applying a scatter filter, after which the ICF remains relatively stable, with a minimum at 20 mm FWHM. This is likely due to significant amounts of scattered counts still being present in the 0 mm image, but not in the others.

\begin{figure}
    \centering
    \includegraphics[width=0.75\linewidth]{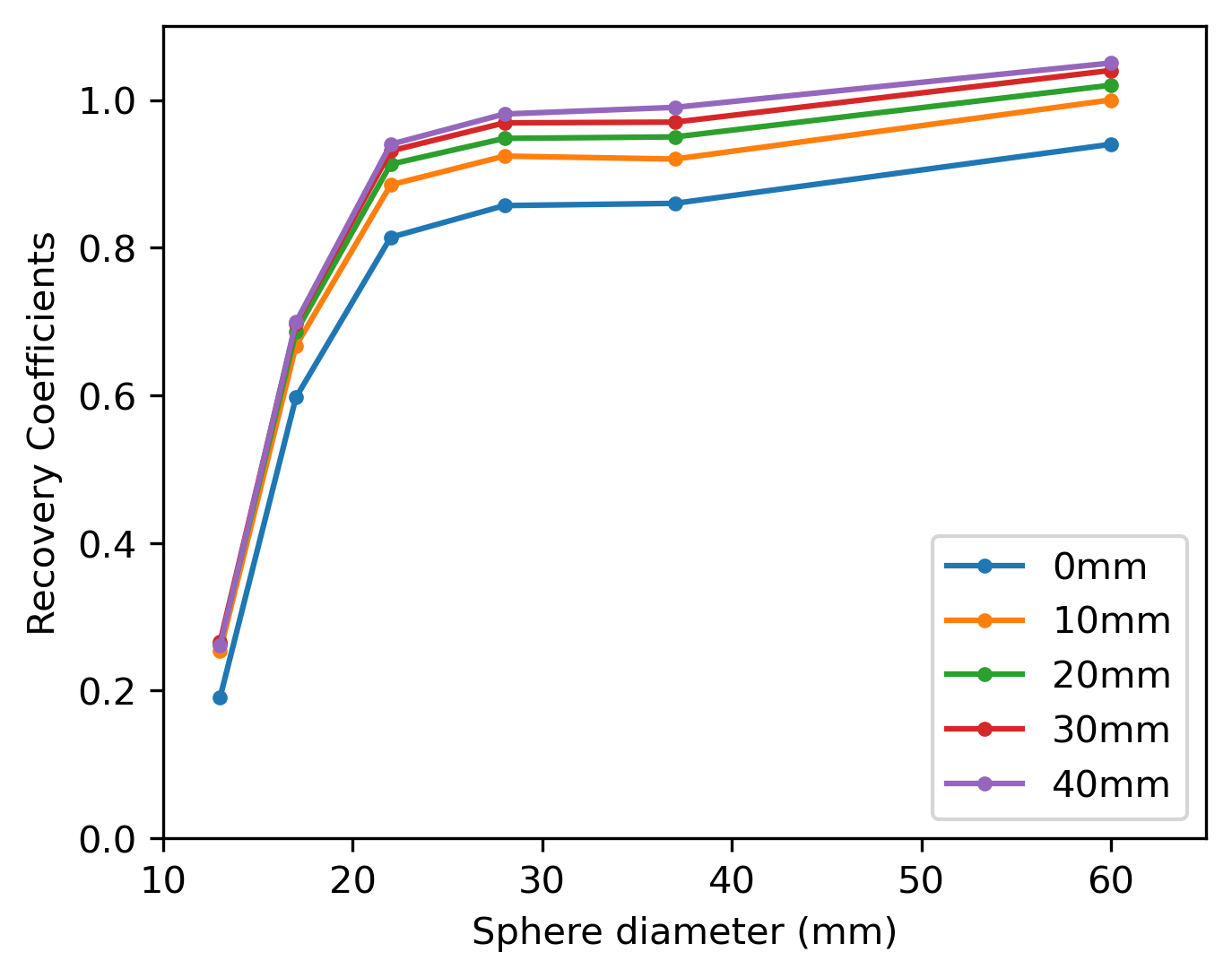}
    \caption{Recovery curves for different levels of scatter window smoothing.}
    \label{fig:SC_smoothing_RCs}
\end{figure}

Scatter window smoothing also has a significant impact on RCs: figure \ref{fig:SC_smoothing_RCs} shows the recovery curves corresponding to the different smoothing levels. RCs increase with increasing FWHM, with the biggest difference between \replaced{10 mm}{10mm} and no smoothing. This is likely due to the contrast-enhancing effect of scatter correction: the scatter information is mostly low-frequency, so correcting for it augments the higher frequencies. Smoothing the scatter estimates only strengthens this effect.

\begin{figure}
    \centering
    \includegraphics[width=0.75\linewidth]{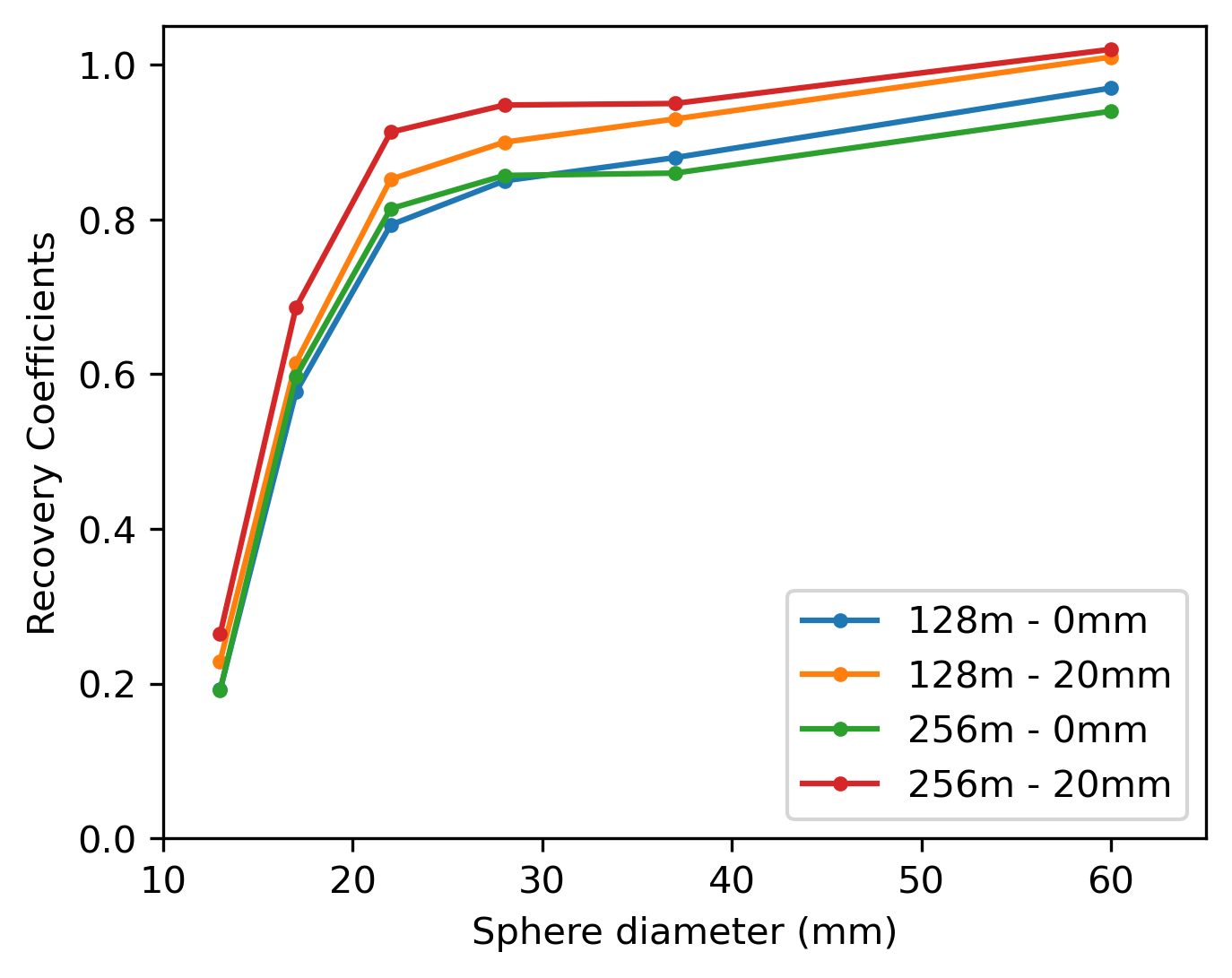}
    \caption{Comparison of RC curves for 128 versus 256 matrices, with and without scatter window smoothing.}
    \label{fig:SC_smoothing_matrix}
\end{figure}

Reducing the matrix size has a similar effect as scatter window smoothing, since it increases the mean counts per pixel, which reduces the noise. In particular, reducing the matrix size from 256 by 256 to 128 by 128 increases the number of counts by a factor of 4, reducing the noise by a factor of 2. As shown in table \ref{tab:ICFs_SC_smoothing}, the 128m and 256m reconstructions with scatter smoothing have the same ICF, but without scatter smoothing, the ICF for the 256m reconstruction is significantly higher. This confirms that reducing the matrix size reduces the bias in the \replaced{image}{scatter estimate}. However, reducing the matrix size also means coarser voxels, which reduces the RCs of small objects, as can be seen in figure \ref{fig:SC_smoothing_matrix}. 

\begin{table}
    \centering
    \caption{Comparison of ICFs (cps/MBq) for different matrix sizes and scatter smoothing levels.}\label{tab:ICFs_SC_smoothing}
    \begin{tabular}{ccc}
    \toprule
        ~ & 0 mm & 20 mm  \\
    \midrule
        $128 \times 128$ & 19.59 & 19.47  \\
        $256 \times 256$ & 20.18 & 19.50  \\
    \bottomrule
    \end{tabular}
\end{table}

These data show that noise in the scatter windows can \deleted{cause biased scatter estimates,} \replaced{affect}{which affects} the reconstructed images both visually and through differences in ICFs and RCs. Reducing the matrix size can improve the scatter estimate, but also affects the rest of the reconstruction by diminishing the information available in the photopeak window. Therefore, it seems preferable to acquire images in a fine matrix and apply a smoothing filter to the scatter windows. Regardless, it is crucial to use a harmonized approach when comparing SPECT images from different systems, since variation in scatter filters can lead to significant differences in the reconstructed images, even when all other reconstruction settings are identical. For this study, 20 mm Gaussian smoothing was found to give the best results, so it was added to the standardized reconstruction settings for the Siemens data. However, the optimal scatter filter likely depends on factors like the activity distribution or the number of acquired counts

\section{Image Calibration Factor Calculation}\label{app:small_versus_large_volume}
A large-volume and a small-volume approach were considered for calculating the ICF, which differed in the volume of interest (VOI) used to derive the total number of counts. The small-volume method, based on \cite{SteffieP}, uses the counts concentration $[C]$ in a VOI that covers only the uniform region of the cylinder, and compares it to the known activity concentration $[A_C]$:

\begin{equation}
    \mathrm{ICF} = \frac{[C]}{\Delta t [A_C]}
\end{equation}
where $\Delta t$ is the acquisition duration.
The large-volume method used in \cite{MRT} employs an expanded VOI around the whole cylinder to capture the total number of counts $C_{\mathrm{tot}}$ originating from the phantom, and compares it to the total activity $A_{\mathrm{tot}}$ 
\begin{equation}
    \mathrm{ICF} = \frac{C_{\mathrm{tot}}}{\Delta t A_{\mathrm{tot}}}
\end{equation}

Both methods were implemented in Python, using a SPECT-based segmentation method. For the large volume method, extending the cylinder ROI by 5 cm in each direction was found to be sufficient to capture (almost) all counts. For the small volume method, average transversal and radial profiles were plotted to determine the size of the uniform region in the image. Based on this, a cylinder of 12 cm diameter and 15 cm length was chosen as it was found to be the largest contour that excluded the Gibbs artifacts in all of the images.

The two methods showed differences of up to 7\%, depending on the camera type and acquisition and reconstruction protocols. In most cases, ICFs determined using the large volume method were 1 to 5\% lower. However, for the Starguide systems, the large volume ICF was up to 3\% higher. Since the cylinder ICFs are used to calculate the RCs for the NEMA phantom, this choice of VOI will also affect the RC calculation. Consistency in the ICF calculation method is therefore crucial when comparing RCs of different systems.

\begin{figure}
    \centering
    \includegraphics[width=0.95\linewidth]{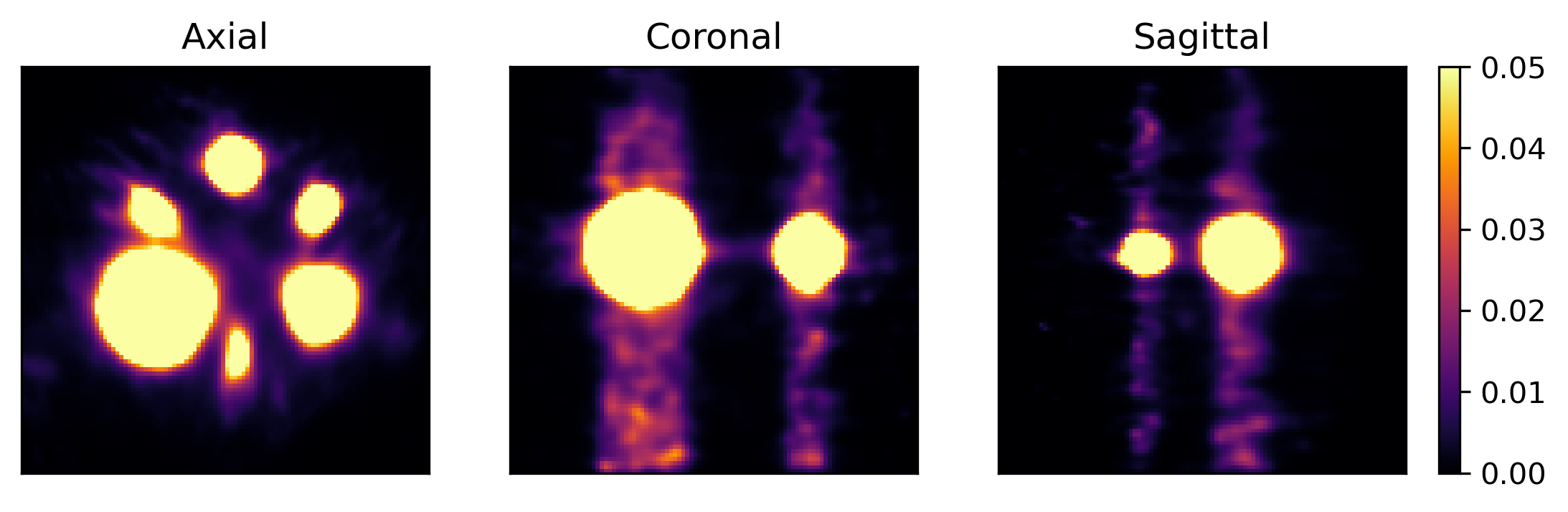}
    \caption{Slices of the NEMA image acquired with the Starguide system B2 using the semi-standardized protocol. Due to septal penetration, a significant number of counts extends beyond the spheres in the axial direction. The data is rescaled such that the expected signal inside the spheres is 1.}
    \label{fig:septal_penetration}
\end{figure}

In this study, the small-volume method was preferred, even though it is harder to implement and slightly less robust than the large-volume method. First, the Starguide images showed a visual septal penetration pattern due to the fixed collimator not being optimized for 208 keV photons (Figure \ref{fig:septal_penetration})\replaced{ so using}{. Using} the large-volume method increases the proportion of septal penetration to true counts \deleted{and thus overestimates the ICF}. Second, the large-volume method includes voxels that contain much more scatter than primary photons, which might result in suboptimal scatter correction when using window-based methods. Finally, the large-volume method includes counts that are incorrectly placed outside of the attenuation map of the phantom (spill-out), therefore leading to incorrect attenuation correction. In contrast, the small-volume method only uses the uniform region of the cylinder, \replaced{which reduces the effect of septal penetration, only includes areas with significant photopeak counts, and avoids spill-out}{where all corrections should perform optimally}. 

\section{Calibration details}\label{app:calibration_details}
\subsection{Radionuclide Calibrators}
The two reference activity meters (a Capintec CRC55tR and a Veenstra VIK-202) at UZ Leuven were calibrated in collaboration with the Laboratory for Nuclear Calibrations (LNK) at the Belgian Study Center for Nuclear Energy (SCK). Their Fidelis, a secondary standard RNC with calibration traceable to primary standards at the National \replaced{Physical}{Physics} laboratory (NPL) in the UK, was set up next to the activity meters of UZ Leuven. Three vials in the reference geometry (10ml Schott vial with 4g of solution) with a nominal activity of 350 MBq, were measured using the Fidelis and the two reference RNCs to determine an average calibration factor for each of the reference RNCs. This factor was used to calculate corrected dial settings based on the instructions in the manual (see below). After changing the dial settings, a new set of measurements was carried out to verify the new calibration.

Since this calibration procedure only used the reference vial, a geometry correction factor for the clinical vial was determined separately. This was done by preparing a uniform stock solution of \lu \, that was used to fill 5 reference vials and 5 clinical vials. The vials were weighed and measured in both RNCs, and the geometry correction factor $f_{\mathrm{geom}}$ was calculated by
\begin{equation}
    f_{\mathrm{geom}} = \frac{\sum_{i=1}^5 {A_{\mathrm{Ref,i}}/m_{\mathrm{Ref,i}}}}{\sum_{i=1}^5 {A_{\mathrm{Clin,i}}/m_{\mathrm{Ref,i}}}}
\end{equation}
Here, $A_{\mathrm{Ref,i}}$ and $m_{\mathrm{Ref,i}}$ are the activity and filling weight of reference vial $i$ and $A_{\mathrm{Clin,i}}$ and $m_{\mathrm{Clin,i}}$ are are the activity and filling weight of clinical vial $i$.

Throughout all measurements, a nominal weight of 4 g and 10 g was assumed for the reference and clinical vials respectively. However, the effective weights of the vials differed up to a few \replaced{decigrams}{decagrams} from the nominal weight. The effect of these deviations was studied by filling three reference and three clinical vials with 1 g of \lu \, solution. The vials were filled by adding small amounts of water and activity measurements with both reference RNCs were performed between each filling step, plotting out the relative response in function of filling volume. The observed volume effect was very small: the response increased by less than 0.1\% per gram for both vial types and both chambers. Therefore, the volume effect was treated as an uncertainty, rather than determining a correction factor for each filling volume.

\paragraph{Determining a new dial setting}
If a traceable reference source is available, it is straightforward to recalibrate a given RNC by calculating a corrected dial setting. In particular, if a source with true activity $A_t$ gives a reading $A_m$, the calibration factor is given by $f = \frac{A_t}{A_m}$. If the initial dial setting is $N_i$, the manual for the CRC-55tR implies that the corrected dial setting $N_c$ is given by
\begin{equation}
    N_c = \frac{1}{f} (N_i + 86.08) - 86.08
\end{equation}
For the VIK-202, the corrected dial setting is given by
\begin{equation}
    N_c = 1000 + \frac{\mathrm{HG}}{10} - f \left[(1000-N_{i})+\frac{\mathrm{HG}}{10}\right]
\end{equation}
This calculation relies on the chamber-specific high gain setting HG, which is set by the vendor upon installation. For the chambers tested in this study, the high gain ranged between 47 and 110.

\paragraph{Uncertainty budget}
Table \ref{tab:unc_RNCs} shows the full uncertainty budgets for the determination of the activity in the two vials using the reference RNCs. The uncertainty on the original calibration was 1.41\%, mostly due to the uncertainty on the reference activity measured by the Fidelis. The two vials used in this study were measured months after the calibration and in a slightly different activity range, so constancy and linearity uncertainties were estimated from existing quality control data. For the latter, it was noted that the response of \isotope[99m]{Tc} was about a factor 3 higher than for \lu \, so the range of interest was shifted accordingly. Next, the statistical uncertainty obtained from the repeated measurements of both vials was added, as well as the uncertainty associated with the deviations from the nominal volume. For the clinical vial, the uncertainty associated with the geometry calculation factor was added as well.

\begin{table}
\centering
\caption{Uncertainty budget for the reference and clinical vials.}\label{tab:unc_RNCs}
\begin{tabular}{lccl}
\toprule
\textbf{Parameter} & \textbf{VIK-202} & \textbf{CRC-55tR} & \textbf{Source} \\
\midrule
Calibration & 1.41\% & 1.41\% & calibration certificate \\
Constancy & 0.3\% & 0.3\% & QC data (\isotope[57]{Co}) \\
Linearity & 0.4\% & 0.4\% & QC data (\isotope[99m]{Tc}) \\

\addlinespace
\multicolumn{4}{l}{\textbf{Reference vial}} \\
Statistical uncertainty & 0.08\% & 0.03\% & repeated measurements \\
Volume effect & 0.03\% & 0.03\% & volume effect calculation \\
\textit{Combined uncertainty} & 1.5\% & 1.5\% &  \\

\addlinespace
\multicolumn{4}{l}{\textbf{Clinical vial}} \\
Statistical uncertainty & 0.05\% & 0.03\% &  repeated measurements \\
Volume effect & 0.03\% & 0.03\% & volume effect calculation \\
Geometry correction & 0.21\% & 0.19\% & geometry correction calculation \\
\textit{Combined uncertainty} & 1.5\% & 1.5\% &  \\
\bottomrule
\end{tabular}
\end{table}

\subsection{Gamma counter}
After the reference RNCs were calibrated with respect to the Fidelis, the gamma counter at UZ Leuven, a Perkin-Elmer Wizard 1480 well counter with a 3 inch NaI(Tl) crystal, was calibrated with respect to the reference RNCs. The measurement protocol of the gamma counter for \lu \, is summarized in table \ref{tab:gammacounter}. For the calibration, 9 tubes containing 1 ml of \lu \, were filled using stock solutions that were prepared from activities measured in the reference RNCs. By performing repeated measurements over several months, the response curve for a wide range of activities could be established. It was found that below 10 kBq, deadtime was negligible and response was linear, with a sensitivity of 39.4 cps/kBq. Therefore, the samples from the NEMA and cylindrical phantoms were left to decay until the activity was below this limit, so the activity concentration could simply be determined by dividing the countrate by the sensitivity and the filling volume. Multiple measurements were performed over several weeks, adapting the measurement time to account for decay, and the results were averaged.

\begin{table}
\centering
\caption{Measurement protocol for the Wizard 1480 gamma counter}\label{tab:gammacounter}
\begin{tabular}{ll}
\toprule
\textbf{Parameter} & \textbf{Value} \\
\midrule
Peak position & 208 keV \\
Window limits & 187.2--228.8 keV \\
Counting Window & Dynamic-keV \\
Corrections & Background, decay \\
Measurement time & Variable\footnotemark[1] \\
\bottomrule
\end{tabular}
\footnotetext[1]{Measurement time was adapted to account for the decay between measurements.}
\end{table}

\paragraph{Uncertainty budget}
Table \ref{tab:unc_phantoms} shows the uncertainty budget for the determination of the activity in the two phantoms using the gamma counter. The uncertainty on the reference activities used to calibrate the gamma counter was assumed to be the same as the combined uncertainty for the two vials described above, since both were measured using the reference RNCs. To this, the uncertainty associated with the calibration itself, i.e. statistical uncertainty and activity concentration variations between the different tubes, was added. Constancy was again estimated from existing QC data, but no separate linearity uncertainty was needed since the measurements were performed in the linear range of the counter. Finally, the statistical uncertainty obtained through repeated measurements and the uncertainty on the phantom volume was added to obtain the full combined uncertainty for the activity in the cylinder. For the NEMA phantom, an extra 1\% uncertainty was added to account for the discrepancy in activity concentration between the samples taken before and after the SPECT measurements.

\begin{table}
\centering
\caption{Uncertainty budget for the uniform and NEMA phantoms.}\label{tab:unc_phantoms}
\begin{tabular}{lcl}
\toprule
\textbf{Parameter} & \textbf{Value} & \textbf{Source} \\
\midrule
Reference activity & 1.5\% & see table \ref{tab:unc_RNCs} \\
Calibration & 0.35\% & calibration measurements \\
Constancy & 0.33\% & QC data (\cs) \\

\addlinespace
\multicolumn{3}{l}{\textbf{Uniform cylinder}} \\
Statistical uncertainty & 0.18\% & repeated measurements \\
Phantom volume & 0.5\% & filling with water \\
\textit{Combined uncertainty cylinder} & \textbf{1.7\%} & \\

\addlinespace
\multicolumn{3}{l}{\textbf{NEMA phantom}} \\
Statistical uncertainty & 0.15\% & repeated measurements \\
Concentration change & 1.0\% & before/after measurements \\
Phantom volume & 0.73\% &  filling with water \\
\textit{Combined uncertainty NEMA} & \textbf{2.0\%} & \\
\bottomrule
\end{tabular}
\end{table}

\section{Supplemental data}\label{app}


\begin{figure}[!h]
    \centering
    \includegraphics[width=0.8\linewidth]{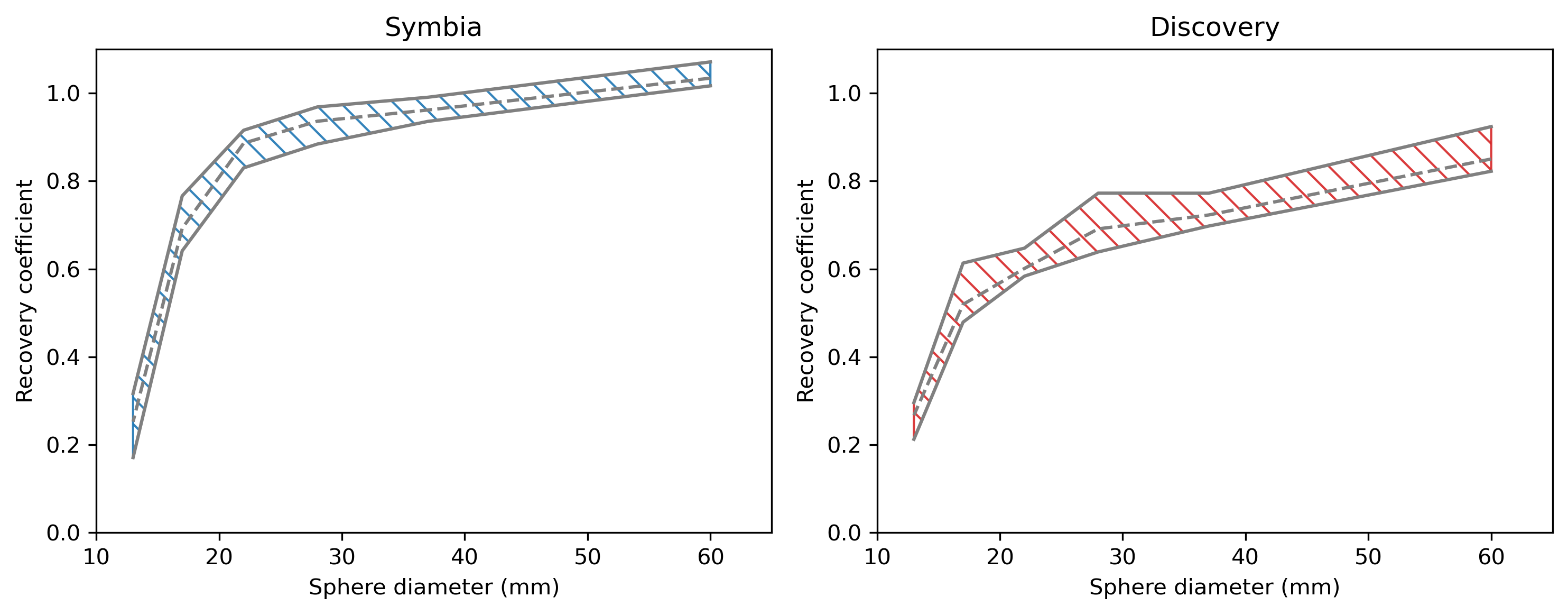}
    \caption{RC bands for the Symbia and Discovery scanners using the fully standardized protocol. The dashed line denotes the mean RC.}
    \label{fig:RC_bands_FS}
\end{figure}

\begin{figure}
    \centering
    \includegraphics[width=0.8\linewidth]{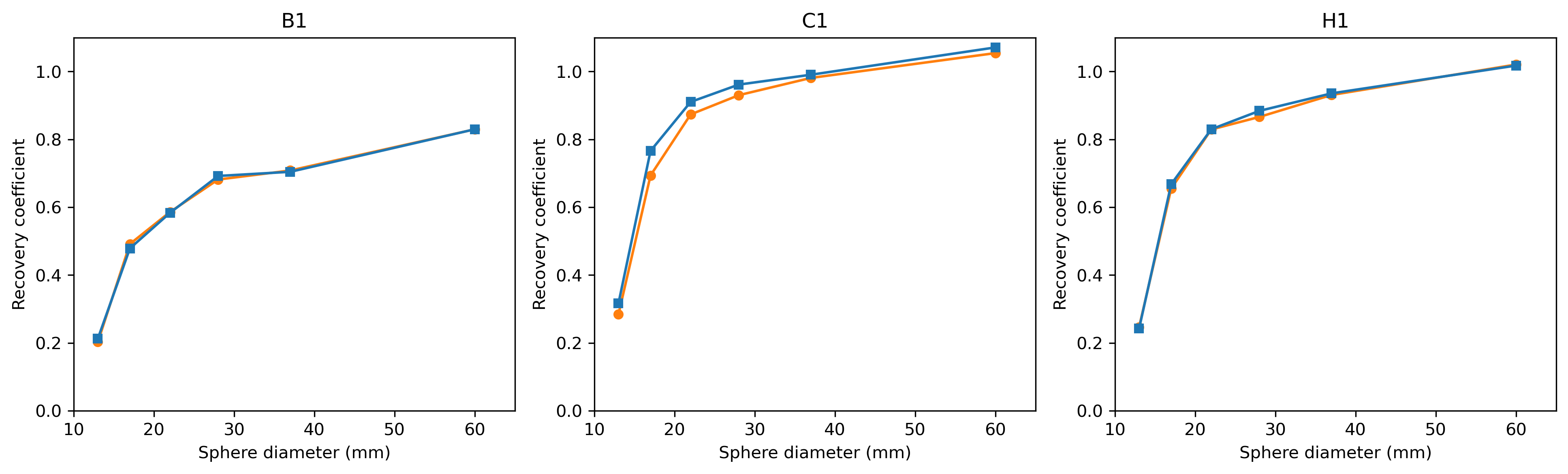}
    \caption{\added{Side-by-side comparison of the recovery curves systems B1, C1 and H1 using fully standardized (blue) versus semi-standardized protocols (orange).}}
    \label{fig:SSvsFSx3}
\end{figure}

\begin{figure}[!h]
    \centering
    \includegraphics[width=\linewidth]{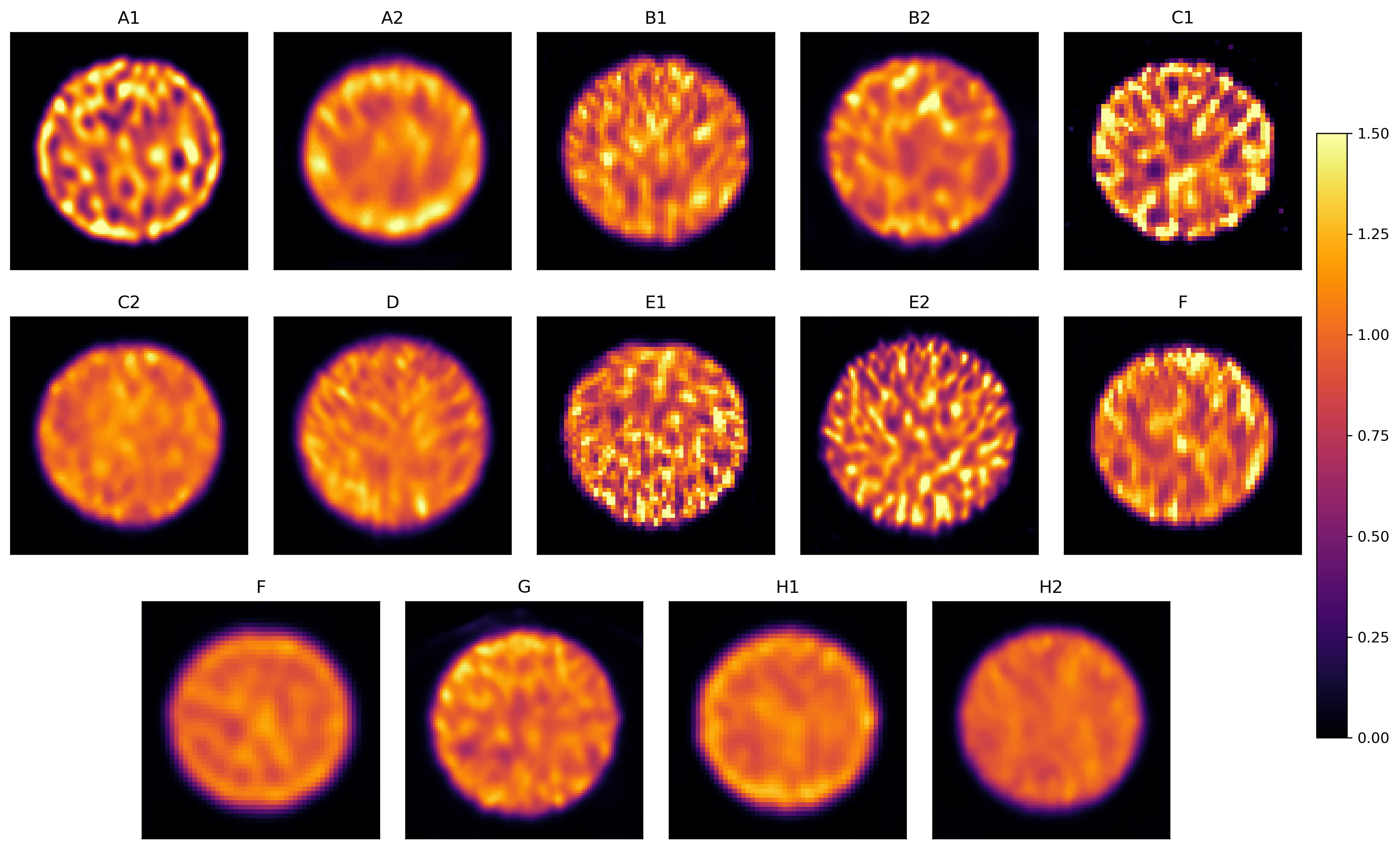}
    \caption{Axial slices of the non-standardized uniform phantom images for the different scanners. \added{Images were normalized such that the expected signal inside the cylinder is equal to 1.}}
    \label{fig:Cylinders_clinical}
\end{figure}

\begin{figure}[!h]
    \centering
    \includegraphics[width=\linewidth]{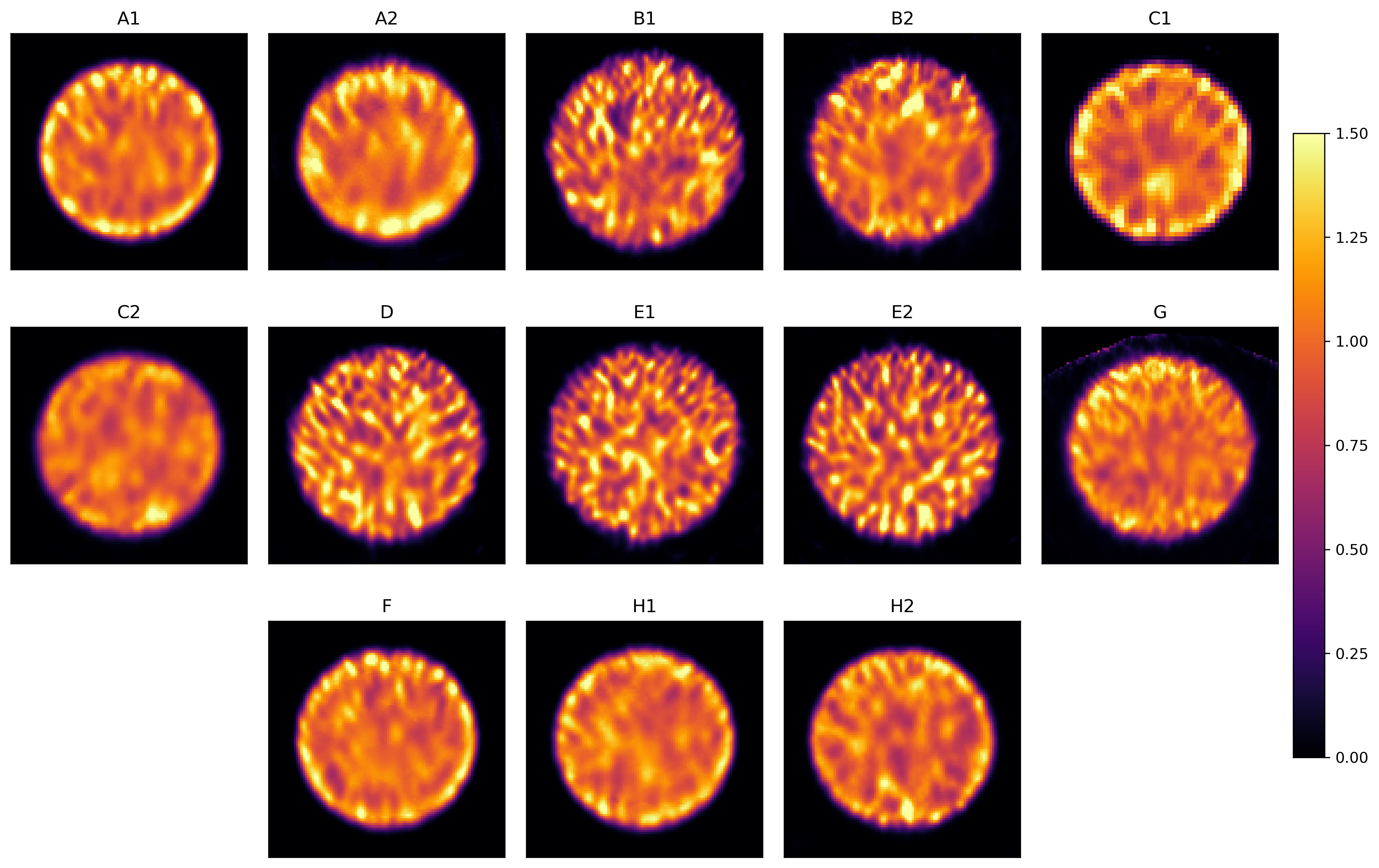}
    \caption{Axial slices of the fully standardized uniform phantom images for the different scanners. \added{Images were normalized such that the expected signal inside the cylinder is equal to 1.}}
    \label{fig:Cylinders_standardized}
\end{figure}

\begin{figure}[!h]
    \centering
    \includegraphics[width=\linewidth]{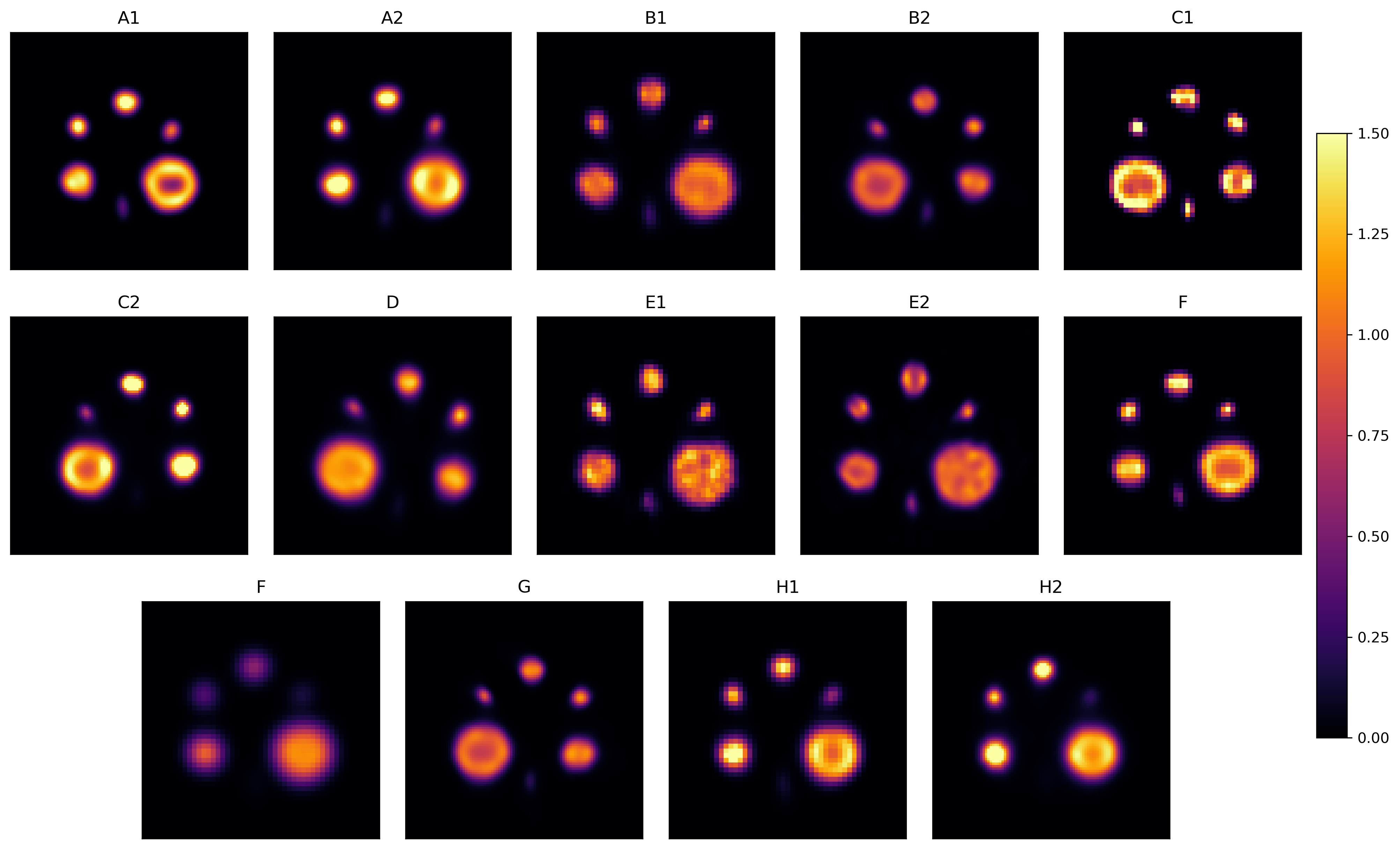}
    \caption{Axial slices of the non-standardized NEMA phantom images for the different scanners. \added{Images were normalized such that the expected signal inside the spheres is equal to 1.}}
    \label{fig:NEMAs_clinical}
\end{figure}

\begin{figure}[!h]
    \centering
    \includegraphics[width=\linewidth]{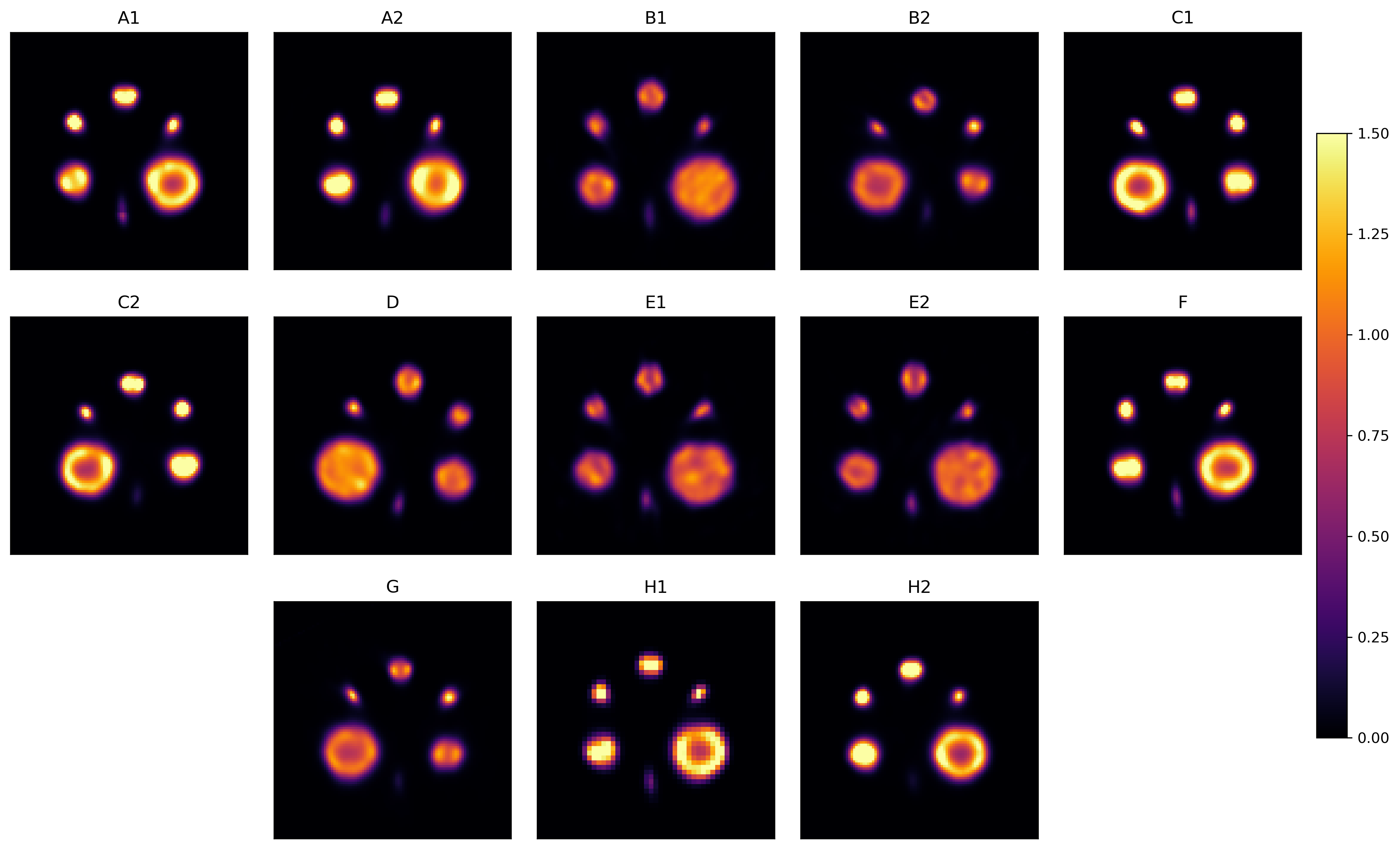}
    \caption{Axial slices of the fully standardized NEMA phantom images for the different scanners. \added{Images were normalized such that the expected signal inside the spheres is equal to 1.}}
    \label{fig:NEMAs_standardized}
\end{figure}

\begin{figure}[!h]
    \centering
    \includegraphics[width=0.75\linewidth]{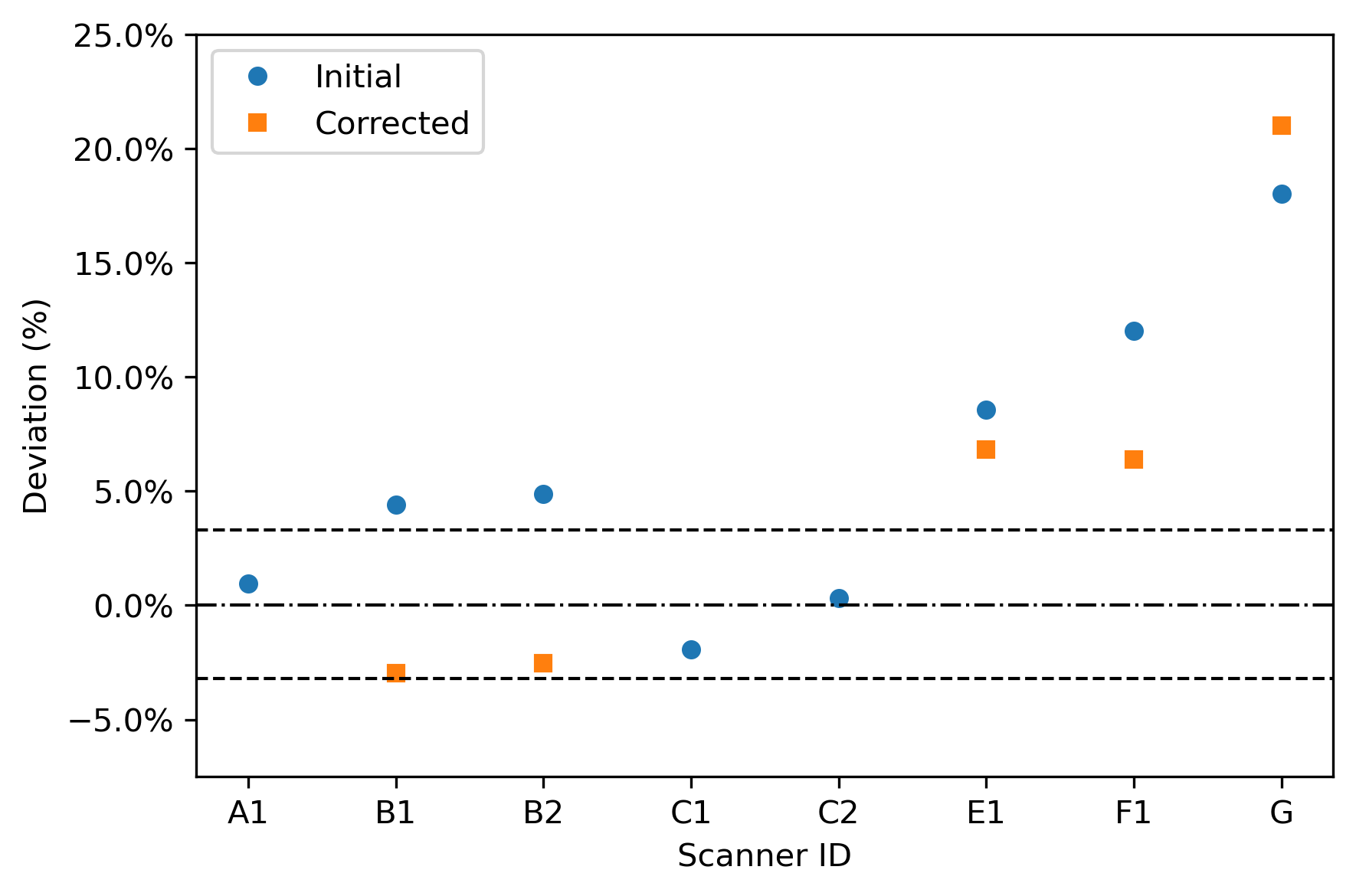}
    \caption{\added{Quantification results for the cylindrical phantom (Figure \ref{fig:SPECT_quant}), corrected for the RNC deviation at the same hospital (Figure \ref{fig:RNCs_clinical}), simulating a scenario where the different RNCs are recalibrated using a traceable reference source. For four scanners, quantification accuracy is improved, bringing two of them within the 95\% confidence interval on the activity in the cylinder. However, deviations remain significant for 3 scanners indicating other calibration uncertainties play a major role. The result for system A1 was not corrected since its calibration does not trace back to the RNC, and no RNC data was available for hospital C. Note that this is a hypothetical exercise; in practice we would recommend recalibrating the SPECT/CT system if the calibration of the reference RNC is changed, rather than applying a correction factor post reconstruction.}}
    \label{fig:Spect_Quant_corrected}
\end{figure}

\begin{table}[!h]
\centering
\begin{tabular}{cc}
\toprule
ID & Geometry correction \\
\midrule
Ref1 & $1.006 \pm 0.002$ \\
A  & $1.006 \pm 0.002$ \\
\multirow{2}{*}{B}  & $1.006 \pm 0.002$\\
~& $1.009\pm0.002$ \\
D  & $1.008 \pm 0.002$ \\
E  & $1.007 \pm 0.002$ \\
G  & $1.006 \pm 0.002$ \\
H1 & $1.006 \pm 0.002$ \\
\midrule
Ref2 & $1.004 \pm 0.002$ \\
F  & $1.005 \pm 0.002$ \\
H2 & $1.004 \pm 0.002$ \\
\bottomrule
\end{tabular}
\caption{Calculated geometry correction factors for the different VIK-202 (top) and CRC-55tR chambers (bottom). For RNC B, different factors are derived for the standard and elevated vial position respectively.}\label{tab:geometry_correction}
\end{table}


\begin{table}[!h]
    \centering
    \caption{Recovery bands for the semi-standardized protocols. Mean, minimum and maximum RCs for each sphere are shown}
    \begin{tabular}{c c c c c c c c c c c c c}
    \toprule
        ~ & \multicolumn{2}{c}{60mm} & \multicolumn{2}{c}{37mm} & \multicolumn{2}{c}{28mm} & \multicolumn{2}{c}{22mm} & \multicolumn{2}{c}{17mm} & \multicolumn{2}{c}{13mm} \\
        ~ & Mean & Range & Mean & Range & Mean & Range & Mean & Range & Mean & Range & Mean & Range \\
    \midrule
        \multirow{2}{*}{Symbia}    
                  & \multirow{2}{*}{1.028} & 1.054 & \multirow{2}{*}{0.953} & 0.988 & \multirow{2}{*}{0.913} & 0.969 & \multirow{2}{*}{0.866} & 0.916 & \multirow{2}{*}{0.662} & 0.694 & \multirow{2}{*}{0.237} & 0.284 \\
                  &  & 1.004 & ~ & 0.919 & ~ & 0.850 & ~ & 0.799 & ~ & 0.634 & ~ & 0.171 \\
        \multirow{2}{*}{Discovery} 
                  & \multirow{2}{*}{0.852} & 0.924 & \multirow{2}{*}{0.726} & 0.772 & \multirow{2}{*}{0.687} & 0.772 & \multirow{2}{*}{0.602} & 0.647 & \multirow{2}{*}{0.526} & 0.613 & \multirow{2}{*}{0.264} & 0.296 \\
                  & ~ & 0.822 & ~ & 0.706 & ~ & 0.639 & ~ & 0.586 & ~ & 0.482 & ~ & 0.203 \\
        \multirow{2}{*}{Starguide} 
                  & \multirow{2}{*}{0.764} & 0.781 & \multirow{2}{*}{0.675} & 0.693 & \multirow{2}{*}{0.671} & 0.686 & \multirow{2}{*}{0.581} & 0.589 & \multirow{2}{*}{0.464} & 0.476 & \multirow{2}{*}{0.110} & 0.123 \\
                  & ~ & 0.747 & ~ & 0.658 & ~ & 0.656 & ~ & 0.573 & ~ & 0.452 & ~ & 0.097 \\
        \multirow{2}{*}{Veriton}   
                  & \multirow{2}{*}{0.997} & 0.998 & \multirow{2}{*}{0.892} & 0.916 & \multirow{2}{*}{0.957} & 0.980 & \multirow{2}{*}{0.921} & 0.955 & \multirow{2}{*}{0.629} & 0.670 & \multirow{2}{*}{0.103} & 0.121 \\
                  & ~ & 0.996 & ~ & 0.868 & ~ & 0.934 & ~ & 0.887 & ~ & 0.587 & ~ & 0.085 \\
    \bottomrule
    \end{tabular}
\end{table}

\begin{table}[!h]
    \centering
    \caption{Recovery bands for the fully standardized protocols. Mean, minimum and maximum RCs for each sphere are shown}
    \begin{tabular}{c c c c c c c c c c c c c}
    \toprule
        ~ & \multicolumn{2}{c}{60mm} & \multicolumn{2}{c}{37mm} & \multicolumn{2}{c}{28mm} & \multicolumn{2}{c}{22mm} & \multicolumn{2}{c}{17mm} & \multicolumn{2}{c}{13mm} \\
        ~ & Mean & Range & Mean & Range & Mean & Range & Mean & Range & Mean & Range & Mean & Range \\
    \midrule
        \multirow{2}{*}{Symbia}    
                  & \multirow{2}{*}{1.034} & 1.071 & \multirow{2}{*}{0.962} & 0.991 & \multirow{2}{*}{0.936} & 0.969 & \multirow{2}{*}{0.886} & 0.916 & \multirow{2}{*}{0.692} & 0.766 & \multirow{2}{*}{0.252} & 0.316 \\
                  & ~ & 1.017 & ~ & 0.936 & ~ & 0.884 & ~ & 0.830 & ~ & 0.641 & ~ & 0.171 \\
        \multirow{2}{*}{Discovery} 
                  & \multirow{2}{*}{0.850} & 0.924 & \multirow{2}{*}{0.723} & 0.772 & \multirow{2}{*}{0.691} & 0.772 & \multirow{2}{*}{0.601} & 0.647 & \multirow{2}{*}{0.519} & 0.613 & \multirow{2}{*}{0.267} & 0.296 \\
                  & ~ & 0.822 & ~ & 0.698 & ~ & 0.639 & ~ & 0.583 & ~ & 0.479 & ~ & 0.212 \\
    \bottomrule
    \end{tabular}
\end{table}

\end{appendices}

\clearpage
\bibliography{sn-bibliography}


\end{document}